\documentclass[sn-mathphys-num]{sn-jnl}
\usepackage{epsfig}%
\usepackage{graphicx}%
\usepackage{multirow}%
\usepackage{amsmath,amssymb,amsfonts}%
\usepackage{amsthm}%
\usepackage{mathrsfs}%
\usepackage{hyperref}
\hypersetup{colorlinks=true,linkcolor=red,citecolor=green,urlcolor=blue}
\usepackage[title]{appendix}%
\usepackage{xcolor}%
\usepackage{textcomp}%
\usepackage{manyfoot}%
\usepackage{booktabs}%
\usepackage{algorithm}%
\usepackage{algorithmicx}%
\usepackage{algpseudocode}%
\usepackage{listings}%
\usepackage{slashed} 
\usepackage{float} 
\begin{document}

\title[Physics on and off the light cone]{Physics on and off the light cone}


\author{\fnm{Philip D.} \sur{Mannheim}}\email{philip.mannheim@uconn.edu}
\affil{\orgdiv{Department of Physics}, \orgname{University of Connecticut},
\orgaddress{ \city{Storrs},   \postcode{ CT 06269}, \country{USA}}}

\abstract
{We study light-front physics and conformal symmetry, and their interplay both on and off the light cone. The full symmetry of the light cone is conformal symmetry not just Lorentz symmetry. Spontaneously breaking conformal symmetry gives masses to particles and takes them off the light cone. Canonical quantization specifies equal-time commutators on the light cone. Equal instant-time and equal light-front-time commutators look very different, but can be shown to be equivalent by looking at unequal-time commutators. We discuss the connection of the light-front approach to the infinite momentum frame approach, and show that vacuum graphs are outside this framework. We show that there is a light-front structure to both AdS/CFT and the eikonal approximation. While mass generation involves scale breaking mass scales, we show that such mass scales can arise via dynamical symmetry breaking in the presence  of scale invariant interactions at a renormalization group fixed point.}

\maketitle

\section{ Minkowski signature predates special relativity}

While Minkowski signature is central to special relativity and light cone studies, it is of interest to note that Minkowski signature predates 20th century special relativity having  originated in differential geometry in the 19th century. To be specific, consider the  2-dimensional Gauss-Bolyai-Lobachevski geometry with line element
\begin{align}
ds^2=\frac{a^2dr^2}{a^2+r^2}+r^2d\theta^2.
\label{LC1.1}
\end{align}
To construct it we introduce a flat 3-dimensional space with a Minkowski-signatured line element
\begin{align}
ds^2=dx^2+dy^2-dt^2,
\label{LC1.2}
\end{align}
as constrained by the hyperbola
\begin{align}
t^2-x^2-y^2=a^2.
\label{LC1.3}
\end{align}
Eliminating $t$ gives
\begin{align}
ds^2=dx^2+dy^2-\frac{(xdx+ydy)^2}{a^2+x^2+y^2}.
\label{LC1.4}
\end{align}
On introducing polar coordinates $x=r\cos\theta$, $y=r\sin\theta$ we recover (\ref{LC1.1}):
\begin{align}
ds^2=dr^2+r^2d\theta^2-\frac{r^2dr^2}{a^2+r^2}=\frac{a^2dr^2}{a^2+r^2}+r^2d\theta^2.
\label{LC1.5}
\end{align}

The significance of the Gauss-Bolyai-Lobachevski geometry is that it did not obey all of Euclid's axioms, to thus open the door to non-Euclidean Riemannian geometry and eventually to General Relativity. It took 2000 years to find  because it does not embed in a Euclidean geometry with line element $ds^2=dx^2+dy^2+dt^2$ but in a geometry with a Minkowski-signatured line element $ds^2=dx^2+dy^2-dt^2$ instead. Technically, the Gauss-Bolyai-Lobachevski geometry is known as a 2-dimensional space of constant negative 2-curvature. Current cosmological studies indicate that we live in a  4-dimensional spacetime with a spatial sector of constant 3-curvature. We will discuss embedding issues again in AdS/CFT.

\section{ Special relativity}
The line element
\begin{align}
ds^2=dt^2-dx^2-dy^2-dz^2=\eta_{\mu\nu}dx^{\mu}dx^{\nu} 
\label{LC1.6}
\end{align}
is Lorentz invariant. It breaks spacetime up into separate timelike ($ds^2>0$), lightlike ($ds^2=0$) and spacelike ($ds^2<0$) regions. Because $ds^2$ is not equal to  the Euclidean-signatured $-dt^2-dx^2-dy^2-dz^2$, one can have nontrivial solutions to $ds^2=0$, viz. the light cone region
where massless particles propagate, with massive particles propagating off the light cone.  To understand the {\bf origin~of~mass} we thus need to understand how to get off the light cone.
To address the origin of mass we need to identify the {\bf full symmetry} of the light cone.

\section{ Conformal symmetry - the full symmetry of the light cone}

While  timelike and spacelike separated  intervals are Lorentz invariant, the light cone itself has a higher symmetry. Its scale invariance is immediate since if  $\eta_{\mu\nu}dx^{\mu}dx^{\nu}=0$ then on scaling $x^{\mu}\rightarrow \lambda x^{\mu}$ we see that $\lambda^2\eta_{\mu\nu}dx^{\mu}dx^{\nu} $ is zero too. 
With 10  constant Poincare parameters $\epsilon^{\mu}$ and  $\Lambda^{\mu}_{\phantom{\mu}\nu}$,  and 5 constant conformal parameters $\lambda$ and $c^{\mu}$ the 15 conformal generators transform $x^{\mu}$ and $x^2$ according to 
\begin{eqnarray}
x^{\mu}&\rightarrow& x^{\mu}+\epsilon^{\mu},\qquad x^{\mu} \rightarrow \Lambda^{\mu}_{\phantom{\mu}\nu}x^{\nu},
\nonumber\\
x^{\mu}&\rightarrow& \lambda x^{\mu},\qquad x^{\mu} \rightarrow \frac{x^{\mu}+c^{\mu}x^2}{1+2c\cdot x+c^2x^2},
\nonumber \\
x^2&\rightarrow& \lambda^2x^2,\qquad x^2 \rightarrow \frac{x^2}{1+2c\cdot x+x^2}.
\label{25}
\end{eqnarray}
The 10 Poincare generators preserve any $x^2$, while the 5 conformal generators also preserve $x^2=0$. Thus in total the $x^2=0$ light cone is preserved by 15 transformations.

\section{The conformal group}

The 15 infinitesimal generators that produce (\ref{25}) act on the coordinates $x^{\mu}$ according to 
\begin{eqnarray}
&&P^{\mu}=i\partial^{\mu},\qquad M^{\mu\nu}=i(x^{\mu}\partial^{\nu}-x^{\nu}\partial^{\mu}),
\qquad
D=ix^{\mu}\partial_{\mu},
\nonumber\\
&&C^{\mu}=i(x^2\eta^{\mu\nu}-2x^{\mu}x^{\nu})\partial_{\nu},
\label{26}
\end{eqnarray}
and together they form the 15-parameter $SO(4,2)$ conformal group:
\begin{eqnarray}
&&[M_{\mu\nu},M_{\rho\sigma}]=i(-\eta_{\mu\rho}M_{\nu\sigma}+\eta_{\nu\rho}M_{\mu\sigma}
-\eta_{\mu\sigma}M_{\rho\nu}+\eta_{\nu\sigma}M_{\rho\mu}),
\nonumber\\
&&[M_{\mu\nu},P_{\sigma}]=i(\eta_{\nu\sigma}P_{\mu}-\eta_{\mu\sigma}P_{\nu}),~~~[P_{\mu},P_{\nu}]=0,~~~
\nonumber\\
&&[M_{\mu\nu},D]=0,\quad [D,P_{\mu}]=-iP_{\mu},~~~[M_{\mu\nu},C_{\sigma}]=i(\eta_{\nu\sigma}C_{\mu}-\eta_{\mu\sigma}C_{\nu}),~~~
\nonumber\\
&&[C_{\mu},P_{\nu}]=2i(\eta_{\mu\nu}D-M_{\mu\nu}),~~~~~[C_{\mu},C_{\nu}]=0,~~~[D,C_{\mu}]=iC_{\mu}.
\label{27}
\end{eqnarray}
Here the four $P_{\mu}$ generate translations, the six antisymmetric  $M_{\mu\nu}$ generate Lorentz transformations, the one $D$ generates scale transformations, and the four $K_{\mu}$ generate what are known as special conformal transformations.

\section{Spinors and conformal symmetry}

The fundamental representation of the conformal group is a 4-dimensional spinor representation since the 15 Dirac matrices $\gamma_5$, $\gamma_{\mu}$, $\gamma_{\mu}\gamma_5$, $[\gamma_{\mu},\gamma_{\nu}]$ also close on the $SO(4,2)$ algebra according to: 
\begin{eqnarray}
 M_{\mu\nu}=\frac{i}{4}[\gamma_{\mu},\gamma_{\nu}], \quad P_{\mu}+C_{\mu}=\gamma_{\mu}\gamma_5,\quad P_{\mu}-C_{\mu}=\gamma_{\mu},\quad D=\frac{i}{2}\gamma_5.
 \label{27a}
 \end{eqnarray}
The group $SU(2,2)$ that contains these spinors is the covering group of $SO(4,2)$ with the 4-dimensional spinor being its fundamental representation.

4-component Dirac spinors are {\bf reducible} under the Lorentz group. They reduce to irreducible  left-handed and right-handed Weyl spinors, viz. the Dirac spinor behaves as  the $D(1/2,0) \oplus D(0,1/2)$ representation. This is puzzling: why should the fundamental building blocks of matter (viz. fermions) be reducible under the fundamental group (viz. the Lorentz group)? Solution: let a bigger group be the fundamental group, one that contains the Lorentz group as a subgroup and under which 4-component Dirac fermions are \textbf{irreducible}.
This is the case for the conformal group, since under it all four Dirac fermion components are {\bf irreducible}, with the conformal transformations mixing the left-handed and right-handed spinors, doing so via  transformations that are continuous.

\section{ Implications of conformal symmetry}

Since conformal symmetry has to hold for all spinors no matter what their internal quantum numbers might be, in a conformal invariant theory  neutrinos would have to have four components too, with {\bf right-handed neutrinos} being needed to accompany the observed left-handed ones. The weak interaction has to be left-right symmetric: $SU(2)_L\times SU(2)_R\times U(1)$, as then generalized to the 3-family $SU(6)_L\times SU(6)_R\times U(1)$. This is Quantum Flavordynamics, in which {\bf all} of the global chiral symmetry of Quantum Chromodynamics is gauged. Why only gauge its $SU(2)_L\times U(1)$ subgroup? -- it is too lopsided. So gauge all of it.

If the conformal symmetry is exact then all particles are massless. Thus we need to generate mass spontaneously. Thus to get off the light cone we need {\bf spontaneous symmetry breaking}. And since the standard Higgs model double-well potential $V(\phi)=\lambda \phi^4-\mu^2\phi^2$ is not conformal invariant because of  its $\mu^2$ mass parameter, the breaking must be done by radiative loops, to hence be {\bf dynamical}.

We must generate masses dynamically for all massive particles, but especially for right-handed neutrinos since their lack of detection to date means that their masses are much larger than those of the left-handed ones. Thus they must acquire Majorana masses, which only involve neutrinos of the same handedness, rather than Dirac masses, which involve both  right-handed and left-handed neutrinos (i.e.,  breaking via a nonzero $\langle\Omega |\psi^C\psi|\Omega\rangle$, rather than a nonzero  $\langle\Omega |\bar{\psi}\psi|\Omega\rangle$  -- see \cite{Mannheim1980}, where some other references to chiral weak interaction studies may also be found). Right-handed neutrino Majorana masses will break parity and reduce the chiral $SU(2)_L\times SU(2)_R\times U(1)$ to $SU(2)_L\times U(1)$ by making right-handed $W$ and $Z$ bosons  heavier than the left-handed ones.

So parity must be broken spontaneously. This resolves a puzzle: If time translations and space reflections commute how could the $[H,P]$ commutator not be zero? Answer: it is zero, but parity is broken in the vacuum, i.e., in the states not in the operators.

If we now make the conformal symmetry local and require invariance under $g_{\mu\nu}(x)\rightarrow e^{2\alpha(x)}g_{\mu\nu}(x)$ with a local $\alpha(x)$, we are led to conformal gravity with action
\begin{align}
I_W=-\alpha_g\int d^4x (-g)^{1/2}C_{\lambda\mu\nu\tau} C^{\lambda\mu\nu\tau},
\label{28a}
\end{align}
where $C_{\lambda\mu\nu\tau}$ is the Weyl tensor and the coupling constant $\alpha_g$ is dimensionless. And we are not led to Einstein gravity with action $I_{EH}=-(c^3/16\pi G_N)\int d^4x(-g)^{1/2}R^{\alpha}_{~\alpha}$,
as it contains Newton's constant $G_N$ with its  intrinsic scale. 

That conformal invariance should be local is motivated by the coupling of the fundamental fermion representation of the conformal group to gravity. While the spin connection was introduced in order to make the Dirac action be locally Lorentz invariant according to
$I_{\rm D}=\int d^4x(-g)^{1/2}i\bar{\psi}\gamma^{c}V^{\mu}_c(\partial_{\mu}+\Gamma_{\mu})\psi$,
where the $V^{\mu}_a$ are vierbeins and $\Gamma_{\mu}=-(1/8)[\gamma_a,\gamma_b](V^b_{\nu}\partial_{\mu}V^{a\nu}+V^b_{\lambda}\Gamma^{\lambda}_{\nu\mu}V^{a\nu})$ is the spin connection, it turns out that this same  Dirac action is locally conformal invariant under $V^{a}_{\mu}(x)\rightarrow e^{\alpha(x)}V^a_{\mu}(x)$, $g_{\mu\nu}(x)\rightarrow e^{2\alpha(x)}g_{\mu\nu}(x)$, $\psi(x)\rightarrow e^{-3\alpha(x)/2}\psi(x)$ with a spacetime-dependent $\alpha(x)$. The spin connection thus acts as a gauge field for local conformal invariance. In addition, we note that 
path integration on $\psi$ and $\bar{\psi}$ for this same Dirac action, viz. $\int D[\psi]D[\bar{\psi}]\exp{iI_{\rm D}}=\exp(iI_{\rm EFF})$, yields an effective action with a  leading term that is none other than the $I_W$ conformal gravity action given in (\ref{28a}) above (see e.g. \cite{'tHooft2010}).

With this thus motivated conformal gravity theory (like Einstein gravity a pure metric theory of gravity that also contains the Schwarzschild solution needed for the solar system) we are able to solve \cite{Mannheim2017} the dark matter, dark energy/cosmological constant and quantum gravity problems, all in one go. Extrapolating Einstein gravity beyond the solar system is where  all the problems come from.

(1) Continuing Einstein gravity to galaxies gives the dark matter problem.

(2) Continuing Einstein gravity to cosmology gives the dark energy/cosmological constant problem.

(3) Quantizing Einstein gravity and continuing the theory far off the mass shell gives the renormalization problem.

(4) Conclusion: With Einstein gravity we might be extrapolating the wrong theory.

The cosmological constant problem is related to mass generation and thus addressed  and, as we show below, solved by the dynamical symmetry breaking mechanism that gets us off the light cone.

\section{Infinite momentum frame}
\begin{figure}[htb]
\begin{center}
\includegraphics[scale=0.6]{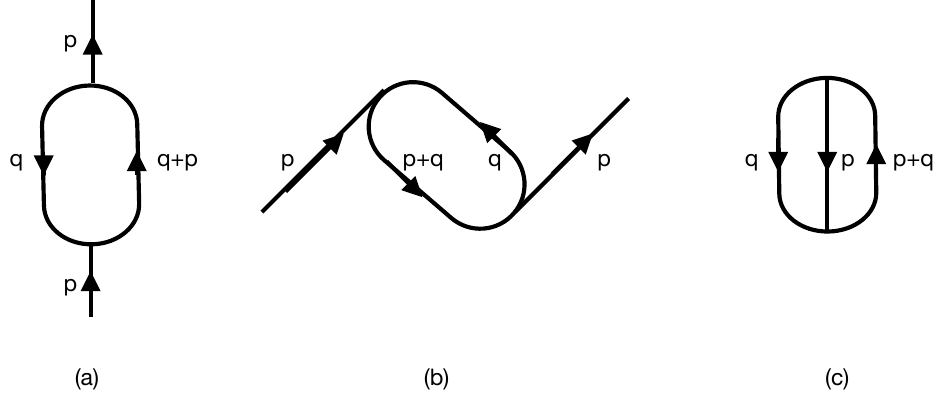}
\caption{Graphs (a), (b) and (c)}
\label{Fig1}
\end{center}
\end{figure}

In 1966 Weinberg \cite{Weinberg1966}  showed that instant-time quantization perturbation theory would be simplified in the frame in which an observer moved with an infinite three-momentum with respect to the center of mass system of a scattering process. Specifically, if we make  a longitudinal Lorentz boost in the $z$ direction with velocity $u$ the momentum components $p^0$ and $p^3$ transform as
\begin{eqnarray}
p^{0\prime}=\frac{(p^0+up^3)}{(1-u^2)^{1/2}},\qquad p^{3\prime}=\frac{(p^3+up^0)}{(1-u^2)^{1/2}}.
\label{K1a}
\end{eqnarray}
For $u$ close to 1, on setting $u=1-\epsilon^2/2$ with $\epsilon$ small,  we obtain 
\begin{eqnarray}
p^{0\prime}=\frac{(p^0+p^3)}{\epsilon},\quad p^{3\prime} =\frac{(p^0+p^3)}{\epsilon}
\label{K1b}
\end{eqnarray}
With $\epsilon$ being small, both $p^{0\prime}$ and $p^{3\prime}$ are very large. Thus on mass shell for a particle of mass $m$ we have 
\begin{align}
p^{0\prime}&=p^{3\prime}\left(1+\frac{((p^1)^2+(p^2)^2+m^2)}{p^{3\prime}}\right)^{1/2} \rightarrow p^{3\prime}+\frac{((p^1)^2+(p^2)^2+m^2)}{2p^{3\prime}}, 
\nonumber\\
p^{0\prime}-p^{3\prime}&=\frac{((p^1)^2+(p^2)^2+m^2)}{2p^{3\prime}}. 
\label{K1d}
\end{align}
Thus $p^{0\prime}$ and $p^{3 \prime}$ are of the same large order of magnitude and $((p^1)^2+(p^2)^2+m^2)/2p^{3\prime}$ is negligible, with the evaluation of Feynman diagrams being simplified. In the infinite momentum frame  Graph (b) is suppressed with respect to Graph (a). Graph (c) was not discussed. In Weinberg's case the $x^0$ time axis runs up the diagram and the analysis was made using old-fashioned perturbation theory. Old-fashioned (i.e.,  pre-Feynman) perturbation theory is off the energy shell but on the mass shell. The Feynman approach itself  is off the mass shell, and when Feynman contours only contain poles the contour integration gives a set of Cauchy residues all of which are on the mass shell. However,  for light-front vacuum tadpole graphs (such as the ones described below)  there  is also a contribution of the contour circle at infinity, and this cannot be described using old-fashioned perturbation theory.

\section{Light-front variables}

In 1969 Chang and Ma \cite{Chang1969}  recast Weinberg's infinite momentum frame analysis in the language of  the light-front variables $p^+=p^0+p^3$, $p^-=p^0-p^3$ that had been  introduced by Dirac \cite{Dirac1949}. Under a longitudinal Lorentz boost in the $z$ direction with velocity $u$ these variables transform as
\begin{eqnarray}
p^{+\prime} =p^+\left(\frac{1+u}{1-u}\right)^{1/2}\rightarrow \frac{p^+}{\epsilon},\quad
p^{-\prime} =p^-\left(\frac{1-u}{1+u}\right)^{1/2}\rightarrow \epsilon p^{-}
\label{K1c}
\end{eqnarray}
With neither of the transverse $p^1$ and $p^2$ components of the 4-momentum $p^{\mu}$ changing under a longitudinal boost, and with $p^+$ and $p^-$ transforming into themselves up to a factor, the light-front structure is maintained under a longitudinal boost.  In terms of light-front variables, Graph (a) as evaluated with a complex plane $p_+$ contour becomes equal to Graph (a) as evaluated with a complex plane $p_0$ contour in an infinite momentum frame with large $p^3$.

For light-front variables we note in general that raising and lowering indices with $p^+=p^0+p^3$, $p^-=p^0-p^3$, $(p^0)^2-(p^3)^2=p^+p^-$,  leads to $p_+=p^-/2$, $p_-=p^+/2$, $p^+p^-=4p_-p_+$, with $p_+$ and $p_-$ respectively being the conjugates of $x^+$ and $x^-$, so that $p\cdot x=p_+x^++p_-x^-+p_1x^1+p_2x^2$.

There is a caveat to the light-front approach. In the infinite momentum frame case the flow of time is forward in $x^0$, while the flow of time in the light-front case is forward in $x^+=x^0+x^3$. But for timelike  events $(x^0)^2 -(x^3)^2=x^+x^-$ is positive, where $x^-=x^0-x^3$. Thus  $x^+x^-$ is positive, and $x^+$ and $x^-$ have the same sign. Thus with the sign of  $x^0=(x^++x^-)/2$  being Lorentz invariant for timelike  events, it follows that when $x^0$ is positive then $x^+$ is positive too for the same timelike events.

 {\bf Thus for  timelike  events, forward in $x^+$ is the same as forward in $x^0$}. 

\section{Relativistic eikonalization and the light-front approach}

In eikonalization of a scalar field $\phi$ with mass $m$ one introduces a phase $\phi=Ae^{iT}$,  and at high momenta with $k_{\mu}k^{\mu}\gg m^2$ one sets $k_{\mu}=\partial T/\partial x^{\mu}=dx_{\mu}/dq$, where $q$ is an affine parameter that measures distance along the eikonal ray. It is immediately suggested to set $T=\int^x k_{\mu}dx^{\mu}$. However, if we do so we would obtain $T=\int^x(dx_{\mu}/dq)dx^{\mu}=\int^x(dx_{\mu}/dq)(dx^{\mu}/dq)dq=\int^xk_{\mu}k^{\mu}dq$, and with $k_{\mu}k^{\mu}=0$, $(dx_{\mu}/dq)(dx^{\mu}/dq)=0$, such a $T$ would vanish identically.

Thus instead we set $T$ equal to the non-vanishing $T=\int^x k_{-}dx^{-}$. Then with $k_+=0$, $k_1=0$, $k_2=0$ one still has $k_{\mu}k^{\mu}=0$  even as $T$ is then nonzero (the vanishing of $k_+$, $k_1$ and $k_2$ does not restrict $k_-$ while still keeping $k_{\mu}k^{\mu}=4k_+k_--k_1^2-k_2^2$ zero). Thus while non-relativistic eikonalization occurs with the normal to the wavefront  being in the $x^3$ direction so that a non-vanishing eikonal phase $T$  is given by $T=\int^x k_3dx^3$, $\partial_3T=k_3=dx_3/dq$, in relativistic eikonalization the normal is in the longitudinal $x^-$ direction, with a non-vanishing eikonal phase $T$ being given by $T=\int^xk_-dx^-$, $\partial_+T=0$, $\partial_-T=k_-=dx_-/dq$.

\section{The takeaway}

In their work Chang and Ma showed that 

(1) for Graph (a) $x^+$ is positive and all the $p_+$ poles have both $p_-$ and $p_+$ positive, 

(2) for Graph (b) $x^+$ is negative and all the $p_+$ poles have both $p_-$ and $p_+$ negative, 

(3) for Graph (c) $x^+$ is zero and so is $p_-$. Thus on shell $p_+=(p_1^2+p_2^2+m^2)/4p_-$ is infinite, just as it should be since it is the conjugate of $x^+$.

However, and this is the key point, \textbf{all of these statements are true without going to the infinite momentum frame.} They thus can define a strategy for evaluating diagrams in which diagrams  are segregated by the sign of the time variable $x^+$. And since   $x^+$ is positive for scattering processes, these processes only involve positive $p_-$ and positive $p_+$, with the $p_+$ pole contributions then corresponding to old-fashioned perturbation theory diagrams. Only needing positive  $p_-$ and $p_+$ provides enormous computational benefits (see e.g. \cite{Leutwyler:1977vy,Burkardt:1995ct,Brodsky:1997de,Bakker:2013cea,deTeramond2013,Brodsky2013,Brodsky2015,deTeramond2024}).  As we discuss below, Refs. \cite{deTeramond2013,Brodsky2013,Brodsky2015,deTeramond2024} are of particular interest to us here as they study the interplay of the light-front approach with conformal symmetry. (Interestingly, the transformations on $p^{+}$ and $p^{-}$ given in (\ref{K1c}) can be thought of as scale transformations.) With exact conformal symmetry requiring that particles be on the light cone, to generate masses and take particles off the light cone we will need spontaneous symmetry breaking, an issue we explore in some detail below.

But what about the instant-time graphs that are not at infinite momentum? Are they different from or the same as the light-front graphs? And if they are different, then which ones describe the real world?  In Mannheim, Lowdon and  Brodsky \cite{Mannheim2021} they were shown to be  the same. Thus Graph (a) in the light-front formalism is equivalent to Graphs (a) and (b) in instant time quantization {\bf at any momenta}. Essentially, in c-number Feynman diagrams the transformation from instant-time coordinates and momenta to light-front coordinates and momenta is just a change of variables.

The tadpole vacuum Graph (c) is expressly non-zero, something known as early as 1969. However, it involves $p_-=0$ zero modes, whose evaluation is tricky. The issue was resolved in Mannheim, Lowdon and Brodsky \cite{Mannheim2019a}. The procedure is to construct the vacuum graph as the $x^+=0$ limit of the light-front time-ordered product  $\langle \Omega |[ \theta(x^+)\phi(x)\phi(0)+\theta(-x^+)\phi(0)\phi(x)]|\Omega\rangle$ (i.e., expressly not the limit of a point-split  2-point function). The $x^+=0$ limit is the limit of two time orderings (forward and backward), even though Graphs (a) and (b) only involve the forward $x^+>0$. Thus the vacuum graph cannot be evaluated using old-fashioned 
3-dimensional on mass shell perturbation theory (though nonvacuum  graphs can be). The vacuum graph must be evaluated as a 4-dimensional off-shell Feynman diagram, which through its circle at infinity contribution contains information that is not accessible using the 3-dimensional approach. With the circle at infinity contribution the light-front tadpole vacuum graph is then explicitly nonzero \cite{Mannheim2019a}.

With the transformation $p^{\pm}=p^0\pm p^3$ only leading to a change of variables in a momentum space Feynman diagram, the value of the Feynman diagram cannot change. However, this only means that the net sum of all pole, cut and circle at infinity contributions to the Feynman contour does not change. It does not require that poles transform into poles, cuts into cuts and circles at infinity into circles at infinity. If there are no cut or circle at infinity contributions then pole contributions transform into pole contributions, to thereby justify the on-shell old-fashioned perturbation theory light-front approach that has been used so effectively. In regard to circle at infinity contributions, we note that even though in renormalizable quantum field theories these contributions are suppressed in both nonvacuum and vacuum instant-time graphs, and even though this suppression holds for nonvacuum light-front graphs as well, it does not hold for vacuum light-front graphs. Thus for vacuum tadpole graphs instant-time poles correspond to light-front pole and circle contributions combined, with the instant-time and light-front vacuum tadpole graphs explicitly being both nonzero and equal to each other \cite{Mannheim2019a}.

\section{Light-front quantization -- the tip of the light cone}

Instead of replacing instant-time momenta by light-front momenta in Feynman diagrams, we can obtain a fully-fledged light-front quantum field theory by  constructing equal $x^+$ commutators rather than equal $x^0$ commutators.

 While scalar field instant-time commutators at equal $x^0$ are of the form
\begin{align}
[\phi(x^0,x^1,x^2,x^3), \partial_0\phi(x^0,y^1,y^2,y^3)]&=i\delta(x^1-y^1)\delta(x^2-y^2)\delta(x^3-y^3),
\nonumber\\
[\phi(x^0,x^1,x^2,x^3), \phi(x^0,y^1,y^2,y^3)]&=0,
\label{LC3.2}
\end{align}
with the second one of these two commutators vanishing, 
scalar field light-front commutators at equal $x^+$ are of the form \cite{Neville1971}
\begin{align}
&[\phi(x^+,x^1,x^2,x^-),\phi(x^+,y^1,y^2,y^-)]=-\frac{i}{4}\epsilon(x^--y^-)\delta(x^1-y^1)\delta(x^2-y^2),
\nonumber\\
&[\phi(x^+,x^1,x^2,x^-),2\partial_-\phi(x^+,y^1,y^2,y^-)]=i\delta(x^1-y^1)\delta(x^2-y^2)\delta(x^--y^-),
\label{3.10}
\end{align}
with the second one of these two commutators not vanishing, 
Thus equal $x^0$ scalar field instant-time commutators and  equal $x^+$ scalar field light-front commutators are markedly different from each other.

Gauge field instant-time commutators at equal $x^0$ are of the form
\begin{align}
[A_{\mu}(x^0,x^1,x^2,x^3),\partial_0A_{\nu}(x^0,y^1,y^2,y^3)]&=-ig_{\mu\nu}\delta(x^1-y^1)\delta(x^2-y^2)\delta(x^3-y^3),
\nonumber\\
[A_{\mu}(x^0,x^1,x^2,x^3),A_{\nu}(x^0,y^1,y^2,y^3)]&=0.
\label{LC3.4}
\end{align}
Using gauge fixing, for gauge field light-front commutators at equal $x^+$ we obtain \cite{Mannheim2021}
\begin{align}
[A_{\mu}(x^+,x^1,x^2,x^-),2\partial_-A_{\nu}(x^+,y^1,y^2,y^-)]&=-ig_{\mu\nu}\delta(x^1-y^1)\delta(x^2-y^2)\delta(x^--y^-),
\nonumber\\
[A_{\mu}(x^+,x^1,x^2,x^-), A_{\nu}(x^+,y^1,y^2,y^-)]&=\frac{i}{4}g_{\mu\nu}\delta(x^1-y^1)\delta(x^2-y^2)\epsilon(x^--y^-).
\label{LC3.5}
\end{align}
Again we see that instant-time commutators and light-front commutators are completely different in structure.

\section{ Instant-time and light-front  anticommutators}

Moreover, this difference is even more pronounced in the fermion case. 
While fermion instant-time anticommutators at equal $x^0$ are of the form
\begin{align}
&\Big{\{}\psi_{\alpha}(x^0,x^1,x^2,x^3),\psi_{\beta}^{\dagger}(x^0,y^1,y^2,y^3)\Big{\}}
\nonumber\\
&=\delta_{\alpha\beta}\delta(x^1-y^1)\delta(x^2-y^2)\delta(x^3-y^3),
\label{LC4.1}
\end{align}
fermion light-front anticommutators at equal $x^+$ are of  the form \cite{Chang1973}
\begin{align}
&\big{\{}[\psi_{(+)}]_{\alpha}(x^+,x^1,x^2,x^-),[\psi_{(+)}^{\dagger}]_{\beta}(x^+,y^1,y^2,y^-)\big{\}}
\nonumber\\
&=\Lambda^+_{\alpha\beta}\delta(x^1-y^1)\delta(x^2-y^2)\delta(x^--y^-),
\label{LC4.2}
\end{align}
where
\begin{align}
\Lambda^{\pm}&=\tfrac{1}{2}(1\pm \gamma^0 \gamma^3),~~\Lambda^{+}+\Lambda^{-}=I,\quad(\Lambda^{+})^2=\Lambda^{+},~~ (\Lambda^{-})^2=\Lambda^{-},~~ \Lambda^{+}\Lambda^{-}=0,~~
\nonumber\\
\gamma^{\pm}&=\gamma^0\pm\gamma^3,~~ (\gamma^{\pm})^2=0,~~~~
\psi_{(\pm)}=\Lambda_{\pm}\psi,\quad [\psi^{\dagger}]_{\pm}=[\psi_{\pm}]^{\dagger}.
\label{LC4.3}
\end{align}
Here $\Lambda^{+}$ and $\Lambda^{-}$ are noninvertible projection operators. The $\psi_{(+)}$ and $\psi_{(-)}$ fields are equally noninvertible, and are respectively known in the literature as good and bad fermions. Not only do the equal instant-time and equal light-front time anticommutators differ by the presence of projection operators in the latter, the bad fermion obeys the nonlocal relation
\begin{align}
&\psi_{(-)}(x^+,x^1,x^2,x^-)
\nonumber\\
&=-\frac{i}{4}\int du^-\epsilon(x^--u^-)[-i\gamma^0(\gamma^1\partial_1+\gamma^2\partial_2)+m\gamma^0]\psi_{(+)}(x^+,x^1,x^2,u^-),
\label{LC4.4}
\end{align}
with $\psi_{(-)}$  thus being a constrained variable and only $\psi_{(+)}$ being dynamical.

In addition, the good-bad and bad-bad light-front anticommutators are given by  \cite{Mannheim2019b,Mannheim2020}
\begin{align}
&\Big{\{}\psi_{a}^{(+)}(x^+,x^1,x^2,x^-),[\psi_{(-)}^{\dagger}]_{b}(x^+,y^1,y^2,y^-)\Big{\}}
\nonumber\\
&=\frac{i}{8}\epsilon(x^--y^-)[i(\gamma^-\gamma^1\partial_1^x+\gamma^-\gamma^2\partial_2^x)-m\gamma^-]_{ab}\delta(x^1-y^1)\delta(x^2-y^2),
\label{LC4.5}
\end{align}
\begin{align}
&\Big{\{}\psi_{a}^{(-)}(x^+,x^1,x^2,x^-),[\psi_{(-)}^{\dagger}]_{b}(x^+,y^1,y^2,y^-)\Big{\}}
=\frac{\Lambda^-_{ab}}{16}\left[-\frac{\partial}{\partial x^1}\frac{\partial}{\partial x^1}-\frac{\partial}{\partial x^2}\frac{\partial}{\partial x^2}+m^2\right]
\nonumber\\
&\times 
\int du^-\epsilon(x^--u^-)\epsilon(y^--u^-)\delta(x^1-y^1)\delta(x^2-y^2).
\label{LC4.6}
\end{align}
These have no instant-time counterparts. {\bf The light-front fermion anticommutators are not only completely different from the instant-time ones, they are even not invertible.}

\section{Why they could in principle be different}

In instant time the light cone is $x_0^2-x_1^2-x_2^2-x_3^2=0$. Thus when $x^0=0$ it follows that $x^1=x^2=x^3=0$, to thus put us at the tip of the light cone.
In the light-front formalism the light cone is $x^+x^--x_1^2-x_2^2=0$. Thus when $x^+=0$ it follows only that $x^1=x^2=0$. However $x^-$ is not constrained, to thus allow for an $\epsilon (x^-)$ term in equal $x^+$ commutators, even though no $\epsilon(x^3)$ type term could appear in the equal $x^0$ commutators.
But does this mean that equal $x^0$ quantization and equal $x^+$ quantization correspond to different physical theories, and if so which one would nature follow? So are they different not just in principle but in practice also? As we now show, despite their very different appearances they are in fact equivalent.

\section{Unequal time commutators and anticommutators}

Following  \cite{Mannheim2019b,Mannheim2020}, we note that according to instant-time quantization the {\bf unequal} time instant-time commutator obtained for a free  massless scalar field from its Fock space expansion is of the form 
\begin{align}
&i\Delta(x-y)=[\phi(x^0,x^1,x^2,x^3), \phi(y^0,y^1,y^2,y^3)]
\nonumber\\
&=
\int \frac{d^3pd^3q}{(2\pi)^3(2p)^{1/2}(2q)^{1/2}}\Big{(}[a(\vec{p}),a^{\dagger}(\vec{q})]e^{-ip\cdot x+iq\cdot y}
+[a^{\dagger}(\vec{p}),a(\vec{q})]e^{ip\cdot x -iq\cdot y}\Big{)}
\nonumber\\
&=\int \frac{d^3p}{(2\pi)^32p}\big{(}e^{-ip\cdot (x-y)}-e^{ip\cdot (x-y)}\big{)}
\nonumber\\
&=-\frac{i}{2\pi}\epsilon(x^0-y^0)\frac{\delta(x^0-y^0-|\vec{x}-\vec{y}|)-\delta(x^0-y^0+|\vec{x}-\vec{y}|)}{2|\vec{x}-\vec{y}|}
\nonumber\\
&=-\frac{i}{2\pi}\epsilon(x^0-y^0)\delta[(x^0-y^0)^2-(x^1-y^1)^2-(x^2-y^2)^2-(x^3-y^3)^2]
\label{LC5.1}
\end{align}
on the light cone.
Since this relation holds at all times, it also holds at equal light-front time. Thus we substitute $x^0=(x^++x^-)/2$, $x^3=(x^+-x^-)/2$, $y^0=(y^++y^-)/2$, $y^3=(y^+-y^-)/2$ and  obtain
\begin{align}
i\Delta(x-y)&=-\frac{i}{2\pi}\epsilon[\tfrac{1}{2}(x^++x^--y^+-y^-)]
\nonumber\\
&\times\delta[(x^+-y^+)(x^--y^-)-(x^1-y^1)^2-(x^2-y^2)^2],
\label{LC5.2}
\end{align}
so that at $x^+=y^+$ we obtain 
\begin{align}
i\Delta(x-y)\big{|}_{x^+=y^+}&=[\phi(x^+,x^1,x^2,x^-), \phi(x^+,y^1,y^2,y^-)]
\nonumber\\
&=-\frac{i}{4}\epsilon(x^--y^-)\delta(x^1-y^1)\delta(x^2-y^2).
\label{LC5.3}
\end{align}
As we see, at $x^+=y^+$ the unequal instant-time commutator is equal to the equal light-front time  commutator

Similarly, with the unequal time massless Abelian gauge field  instant-time commutator being of the form  
\begin{align}
&[A_{\mu}(x^0,x^1,x^2,x^3),A_{\nu}(y^0,y^1,y^2,y^3)]=-ig_{\mu\nu}\Delta(x-y)
\nonumber\\
&=\frac{i}{2\pi}g_{\mu\nu}\epsilon(x^0-y^0)\delta[(x^0-y^0)^2-(x^1-y^1)^2-(x^2-y^2)^2-(x^3-y^3)^2],
\label{LC5.4}
\end{align}
 we obtain 
\begin{align}
&[A_{\mu}(x^+,x^1,x^2,x^-), \partial_{-}A_{\nu}(x^+,y^1,y^2,y^-)]=-\frac{i}{2}g_{\mu\nu}\delta(x^1-y^1)\delta(x^2-y^2)\delta(x^--y^-),
\nonumber\\
&[A_{\mu}(x^+,x^1,x^2,x^-), A_{\nu}(x^+,y^1,y^2,y^-)]=\frac{i}{4}g_{\mu\nu}\delta(x^1-y^1)\delta(x^2-y^2)\epsilon(x^--y^-),
\label{LC5.5}
\end{align}
at $x^+=y^+$,
so that the  $x^+=y^+$ unequal instant-time commutator is equal to the equal light-front time  commutator. Similar result holds for non-Abelian gauge fields. 

\textbf{Thus light-front quantization is instant-time quantization, and does not need to be independently postulated.}

\section{Fermion unequal time anticommutators}
When massless fermions are quantized according to standard equal instant-time quantization,  the unequal instant-time anticommutator is given by 
\begin{eqnarray}
\big{\{}\psi_{\alpha}(x^0,x^1,x^2,x^3), \psi_{\beta}^{\dagger}(y^0,y^1,y^2,y^3)\big{\}}=\left[(i\gamma^{\mu}\gamma^0\partial_{\mu}\right]_{\alpha\beta}i\Delta(x-y),
\label{LC6.1}
\end{eqnarray}
where $\Delta(x-y)$ is given in (\ref{LC5.1}). As constructed, (\ref{LC6.1}) involves no $\Lambda^{\pm}$ projectors.  
On now applying the $\Lambda^+$ projection operator, then following some algebra  we obtain 
\begin{align}
&\Lambda^+_{\alpha\gamma}\big{\{}\psi_{\gamma}(x^+,x^1,x^2,x^-),\psi_{\delta}(x^+,y^1,y^2,y^-)\big{\}}\Lambda^+_{\delta\beta}
\nonumber\\
&=\big{\{}[\psi_{(+)}(x^+,x^1,x^2,x^-)]_{\alpha},[\psi_{(+)}^{\dagger}]_{\beta}(x^+,y^1,y^2,y^-)\big{\}}
\nonumber\\
&=\Lambda^+_{\alpha\beta}\delta(x^--y^-)\delta(x^1-y^1)\delta(x^2-y^2).
\label{LC6.2}
\end{align}
at $x^+=y^+$.
We thus recognize the  $x^+=y^+$ unequal instant-time anticommutator as being equal to the equal light-front time good-good anticommutator. We can also derive  anticommutators involving the bad fermions in the same way \cite{Mannheim2019b,Mannheim2020}. The instant time and light front equivalence for bosons thus applies to fermions as well.

\section{The takeaway}

Light-front quantization is instant-time quantization, and does not need to be independently postulated. The seemingly different structure between equal instant-time and equal light-front time commutators is entirely a consequence  of unequal  time commutators and anticommutators being restricted to equal $x^0$ or equal $x^+$. 

Now the transformation $x^{\pm}=x^0\pm x^3$ is not a Lorentz transformation but a translation, i.e., a general coordinate transformation. But for theories that are Poincare invariant this is a symmetry. Thus  any two directions of quantization that can be connected by a general coordinate transformation must describe the same theory. But in the quantum theory translations are unitary transformations. Thus  following  \cite{Mannheim2020,Mannheim2021} we now show that instant-time quantized and light-front quantized theories are unitarily equivalent, to thus indeed  be different manifestations of one and the same theory.

\section{Unitary equivalence via translation invariance}

So far we have only discussed free theory operator commutators, and while they involve quantum operators they themselves just happen to be c-numbers. To generalize the discussion to interacting theories we will instead need to discuss matrix elements of operators, with these matrix elements also being c-numbers.  With Poincare invariance requiring  that the momentum generators obey
\begin{align}
[\hat{P}_{\mu},\phi]=-i\partial_{\mu}\phi,\quad [\hat{P}_{\mu},\hat{P}_{\nu}]=0
\label{LC10.1}
\end{align}
to all orders in perturbation theory, we introduce the unitary  operator 
\begin{eqnarray}
U(\hat{P}_0,\hat{P}_3)=\exp(ix^3\hat{P}_0)\exp(ix^0\hat{P}_3).
\label{LC10.2}
\end{eqnarray}
It  transforms the instant-time (IT) coordinates of a field to the light-front (LF) coordinates of a field according  to
\begin{align}
U\phi(IT;x^0,x^1,x^2,-x^3)U^{-1}&= \phi(IT;x^0+x^3,x^1,x^2,x^0-x^3)
=\phi(LF;x^+,x^1,x^2,x^-).
\nonumber\\
\label{LC10.3}
\end{align}
Then with a light-front vacuum of the form $|\Omega_F\rangle=U|\Omega_I\rangle$ we obtain
\begin{align}
&-i\langle \Omega_I|[\phi(IT;x^0,x^1,x^2,-x^3),\phi(0)]|\Omega_I\rangle
\nonumber\\
&=-i\langle \Omega_I|U^{\dagger}U[\phi(IT;x^0,x^1,x^2,-x^3),\phi(0)]U^{\dagger}U|\Omega_I\rangle
\nonumber\\
&=-i\langle \Omega_F|[\phi(LF;x^+,x^1,x^2,x^-),\phi(0)]|\Omega_F\rangle,
\label{C.10.4}
\end{align}
to all orders in perturbation theory. We thus establish the unitary equivalence of matrix elements of instant-time commutators and light-front commutators to all orders. With the Lehmann representation incorporating these matrix elements, this same equivalence also holds for the all-order Lehmann representation \cite{Mannheim2020,Mannheim2021}.

It is also of interest to relate the eigenvalues of the instant-time and light-front momentum generators. To this end we note that $p^0=(p^++p^-)/2$, so that with $p_-=p^+/2$, $p_+=p^-/2$ we have $p_0=p_-+p_+$.
With the all-order momentum operators having real and complete eigenspectra we have the all-order
\begin{align}
\hat{P}_{\mu}(IT)=\sum|p^n(IT)\rangle p^n_{\mu}(IT)\langle p^n(IT)|,\quad \hat{P}_{\mu}(LF)=\sum|p^n(LF)\rangle p^n_{\mu}(LF)\langle p^n(LF)|.
 \label{LC10.11}
 \end{align}
On applying the unitary transformation $U(\hat{P}_0,\hat{P}_3)$ and recalling  that $[P^{\mu},P^{\nu}]=0$, we obtain
\begin{align}
\hat{P}_0(IT)&=U\hat{P}_0(IT)U^{-1}=U\sum| p^n(IT)\rangle p^n_{0}\langle p^n(IT)|U^{\dagger}
\nonumber\\
&=\sum|p^n(LF)\rangle (p^n_++p^n_-)\langle p^n(LF)|=\hat{P}_+(LF)+\hat{P}_-(LF).
 \label{LC10.12}
 \end{align}

\smallskip

With eigenvalues not changing under a unitary transformation there initially appears to be a mismatch between the eigenvalues of $\hat{P}_0(IT)$ and $\hat{P}_+(LF)$. However, for any timelike set of instant-time momentum eigenvalues we can Lorentz boost $p_1$, $p_2$ and $p_3$ to zero, to yield 
\begin{align}
p_1=0, \quad p_2=0, \quad p_3=0,\quad p_0=m. 
\label{LC.10.13}
\end{align}
If we impose this same $p_1=0$, $p_2=0$, $p_3=0$ condition on the light-front momentum eigenvalues we would set $p_+=p_-$, $p^2=4p_+^2=m^2$, and thus obtain 
\begin{align}
p_1=0, \quad p_2=0, \quad p_+=p_-,\quad p_0=2p_+=m.
\label{LC.10.14}
\end{align}
When written in terms of contravariant vectors with $p^{\mu}=g^{\mu\nu}p_{\nu}$ this condition takes the form 
\begin{align}
p^0=p^-=m.
\label{LC10.15}
\end{align}
Thus in \textbf{the instant-time rest frame} the eigenvalues of the contravariant $\hat{P}^0(IT)$ and $\hat{P}^-(LF)$ coincide. In this sense then instant-time and light-front Hamiltonians are equivalent. \textbf{And thus non-relativistic in the light-front case still means $p_3=0$, i.e., $p_+=p_-$, and not $p_-=p^+/2=0$.}

\section{AdS/CFT}

To study embeddings in $AdS_5$ we follow the work of  Guth, Kaiser, Mannheim and Nayeri, as reported in \cite{Mannheim2005}.

\medskip

An $AdS_5$ geometry can be described by the flat $M(4,2)$ metric 
\begin{equation}
ds^2=dU^2+dV^2-dW^2-dX^2-dY^2-dZ^2,
\label{10.1}
\end{equation}
as subject to the constraint 
\begin{equation}
U^2+V^2-W^2-X^2-Y^2-Z^2=\ell^2.
\label{10.2}
\end{equation}
On introducing 
\begin{align}
U=&te^{-w/\ell},\quad V+W=\ell e^{-w/\ell},\quad  V-W=\ell e^{w/\ell}+\frac{r^2-t^2}{\ell}e^{-w/\ell},
\nonumber\\
X=&re^{-w/\ell}\sin\theta\cos\phi,\quad Y=re^{-w/\ell}\sin\theta\sin\phi,\quad Z=re^{-w/\ell}\cos\theta,
\label{10.47}
\end{align}
we obtain 
\begin{eqnarray}
ds^2=&&-dw^2+e^{-2w/\ell}\left[dt^2-dr^2 -r^2d\theta^2-
r^2\sin^2\theta d\phi^2\right]~~,
\label{10.49}
\end{eqnarray}
to thus embed a 4-dimensional Minkowski brane in a 5-dimensional $AdS_5$ bulk. We thus parallel  the embedding of a 4-dimensional conformal field theory (CFT)  in a 5-dimensional $AdS_5$ bulk. For our purposes here we note that the embedding has to be done with light-front variables $V\pm W$. If we set $du=dwe^{w/\ell}$, $u=\ell e^{w/\ell}$  and switch the Minkowski coordinates  from polar to Cartesian, we can write (\ref{10.49}) in the form
\begin{eqnarray}
ds^2=&&\frac{\ell^2}{u^2}\left[-du^2+dt^2-dx^2-dy^2-dz^2\right],
\label{10.49a}
\end{eqnarray}
which shows that the $AdS_5$ geometry is conformal to a flat 5-dimensional Minkowski metric, with the 5-dimensional Weyl tensor vanishing.

Interestingly, we note that a connection between AdS/CFT and the light-front approach has also been discussed in \cite{deTeramond2013,Brodsky2013,Brodsky2015,deTeramond2024}. Specifically, in \cite{deTeramond2013,Brodsky2013,Brodsky2015,deTeramond2024} the action for a  scalar field $S$ with mass $m$ propagating in this $AdS_5$ background  of the form
\begin{eqnarray}
I_S=&&\frac{1}{2}\int dxdydtdz du(-g)^{1/2}\left[g^{MN}\partial_{M}S\partial_{N}S-m^2S^2\right]
\label{10.49b}
\end{eqnarray}
was introduced, where $M$ and $N$ range over the five coordinates in the  $AdS_5$ space, and $g=-(\ell^2/u^2)^{5}$ is the determinant of the metric given in (\ref{10.49a}).
In order to generate a scale (one that is to become a QCD scale) the action $I_S$ was modified to the form 
\begin{eqnarray}
I_S=&&\frac{1}{2}\int d^5x(-g)^{1/2}e^{\phi(u)}\left[g^{MN}\partial_{M}S\partial_{N}S-m^2S^2\right]
\nonumber\\
=&&\frac{1}{2}\int d^5x\frac{\ell^5}{u^5}e^{\phi(u)}\bigg{[}\frac{u^2}{\ell^2}\left(-\partial_{u}S\partial_{u}S+\partial_{t}S\partial_{t}S-\partial_{x}S\partial_{x}S-\partial_{y}S\partial_{y}S-\partial_{z}S\partial_{z}S\right)
\nonumber\\
&&-m^2S^2\bigg{]}
\label{10.49bb}
\end{eqnarray}
by the introduction of the $e^{\phi(u)}$ factor, with $\phi(u)=\lambda u^2$ with scale parameter $\lambda$  being found to be a particularly appropriate choice.
The wave equation that results from this modified action as evaluated at $u=0$ was then found to be equivalent to the light-front equation for QCD singlet hadron states in the $x,y,z,t$ space, to thus give a modified AdS/CFT correspondence, one explicitly using light-front quantization.

To see in exactly which particular  way the $AdS_5$ geometry has been modified, we take the modified metric to be of the form 
\begin{eqnarray}
ds^2=&&A^2(u)du^2+B^2(u)[dx^2+dy^2+dz^2-dt^2],
\label{10.49c}
\end{eqnarray}
with determinant $-A^2B^8$.
In such a geometry the scalar field action takes the form
\begin{eqnarray}
I_S=&&\frac{1}{2}\int d^5xAB^4\bigg{[}-A^{-2}\partial_{u}S\partial_{u}S 
\nonumber\\
&&+B^{-2}\left[\partial_{t}S\partial_{t}S-\partial_{x}S\partial_{x}S-\partial_{y}S\partial_{y}S-\partial_{z}S\partial_{z}S \right]-m^{\prime 2}S^2\bigg{]},
\label{10.49d}
\end{eqnarray}
where, as we explain momentarily, we have changed the mass parameter from $m$ to $m^{\prime}$. Comparing (\ref{10.49bb}) with (\ref{10.49d}) we obtain
\begin{eqnarray}
\frac{B^4}{A}=\frac{\ell^3}{u^3}e^{\phi},\qquad AB^2=\frac{\ell^3}{u^3}e^{\phi},\qquad AB^4m^{\prime 2}=\frac{\ell^5}{u^5}e^{\phi}m^2.
\label{10.49e}
\end{eqnarray}
From these relations it follows that 
\begin{eqnarray}
A=B=\frac{\ell}{u}e^{\phi/3},\qquad m^{\prime 2}=e^{-2\phi/3}m^2.
\label{10.49f}
\end{eqnarray}
Thus the line element in (\ref{10.49c}) and $I_S$ as given in (\ref{10.49d}) take the form
\begin{eqnarray}
ds^2=&&e^{2\phi/3}\frac{\ell^2}{u^2}\left[-du^2+dt^2-dx^2-dy^2-dz^2\right], \quad
\nonumber\\
I_S=&&\frac{1}{2}\int d^5x\frac{\ell^5 e^{5\phi/3}}{u^5}\bigg{[}\frac{u^2e^{-2\phi/3}}{\ell^2}\big{[}-\partial_{u}S\partial_{u}S +\partial_{t}S\partial_{t}S
\nonumber\\
&&-\partial_{x}S\partial_{x}S-\partial_{y}S\partial_{y}S-\partial_{z}S\partial_{z}S \big{]}-m^{\prime 2}S^2\bigg{]}.
\label{10.49g}
\end{eqnarray}
While no longer an $AdS_5$ metric, this metric is conformal to an $AdS_5$ metric, so for it  the Weyl tensor still vanishes. We note that the relation between $m$ and $m^{\prime}$ is  dependent on $u$.  In order to be able to finish up with a spacetime-independent mass parameter $m^{\prime}$ in (\ref{10.49g}) we must replace $m$ by $me^{\phi/3}$ in the starting (\ref{10.49b}), so that the $I_S$  that is to actually be used for hadron spectra (viz. (\ref{10.49g})) has a spacetime-independent mass parameter.  While the reasoning is different, the need to make the starting $m^2$ spacetime dependent had also been noted in \cite{deTeramond2013}.

We should also note that while the above light-front $AdS/CFT$ approach relies on conformal symmetry, in one respect the approach  is not conformal invariant, namely in the appearance of the mass term in (\ref{10.49b}). To achieve 5-dimensional local conformal invariance under $g_{MN}(x)\rightarrow e^{2\alpha(x)}g_{MN}(x)$, $S(x)\rightarrow e^{-3\alpha(x)/2}S(x)$ ($S(x)\rightarrow e^{(2-D)\alpha(x)/2}S(x)$ in dimension $D$) for arbitrary spacetime-dependent $\alpha(x)$, we would need to drop the $m^2$ term and replace  (\ref{10.49b}) by
\begin{eqnarray}
I_S=&&\frac{1}{2}\int d^5x(-g)^{1/2}\left[g^{MN}\partial_{M}S\partial_{N}S-\frac{3}{16}R^{\alpha}_{~\alpha}S^2\right],
\label{10.49h}
\end{eqnarray}
where $R^{\alpha}_{~\alpha}$ is the Ricci scalar. In $AdS_5$ the Ricci scalar is given by $R^{\alpha}_{~\alpha}=-20/\ell^2$. While this indeed would yield a spacetime-independent mass term,  and indeed would get us off the light cone, unfortunately the mass term is of the form $m^2=-15/4\ell^2$, to thus be tachyonic rather than having a positive mass squared. However, as we now discuss in reference to dynamical mass generation, having a tachyon could indicate that we are in the wrong vacuum, and that there could be a different (spontaneously broken) vacuum in which squared masses are positive. While  it is beyond the scope of this paper to seek any such vacuum in the $AdS_5/CFT$ case, we turn now to a four-fermion model to see how to see how things work in that particular case.

\section{Nambu-Jona-Lasinio chiral four-fermion model as a mean-field theory}

To get off the light cone where there  are no mass scales, we need to generate mass scales dynamically. In order to see how to do this, we first look at the nonrenormalizable four-fermion Nambu-Jona-Lasinio (NJL) model \cite{Nambu1961}  of mass generation via dynamical symmetry breaking, even though this model does have a cut off with an intrinsic mass scale. We then rectify this by dressing the NJL model with an Abelian QED gauge theory that is taken to have a renormalization group fixed point, with its associated anomalous dimensions then enabling us to obtain a renormalizable NJL model. Thus we use scale invariance with anomalous dimensions to get us off the scale invariant light cone. The discussion of the NJL model that we present here follows \cite{Nambu1961,Mannheim1976}.

The NJL model is a chiral-invariant massless four-fermion theory with action 
\begin{eqnarray}
I_{\rm NJL}&=&\int d^4x \left[i\bar{\psi}\gamma^{\mu}\partial_{\mu}\psi-\frac{g}{2}[\bar{\psi}\psi]^2-\frac{g}{2}[\bar{\psi}i\gamma_5\psi]^2\right].
\label{V4}
\end{eqnarray}
We introduce a mass term  $m$ as a trial parameter and rewrite the action as
\begin{eqnarray}
I_{\rm NJL}
&=&\int d^4x \left[i \bar{\psi}\gamma^{\mu}\partial_{\mu}\psi-m\bar{\psi}\psi +\frac{m^2}{2g}\right]
+\int d^4x \left[-\frac{g}{2}\left(\bar{\psi}\psi-\frac{m}{g}\right)^2-\frac{g}{2}\left(\bar{\psi}i\gamma_5\psi\right)^2\right]
\nonumber\\
&=&I_{\rm MF}+I_{\rm RI},
\label{V5}
\end{eqnarray}
i.e., we break up the action into mean-field  and residual-interaction components. We note the presence of the $m^2/2g$ term. It serves as cosmological constant term that will enable us to control the vacuum energy.

\begin{figure}[htb]
\begin{center}
\includegraphics[scale=0.15]{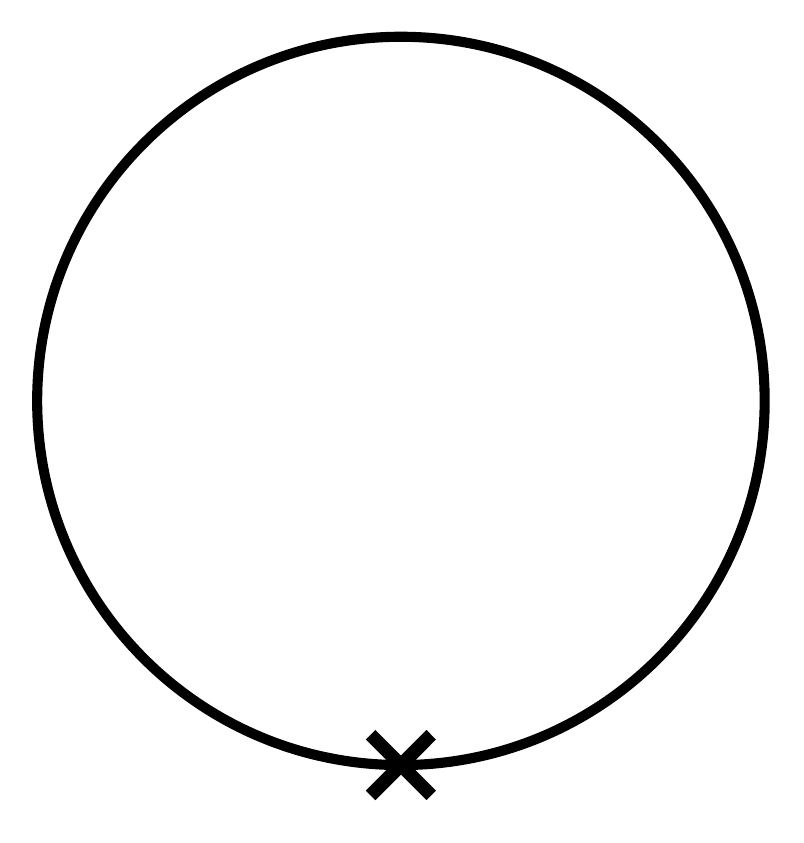}
\caption{The NJL bare vertex vacuum tadpole mass insertion graph for $\langle \Omega_{\rm m}|\bar{\psi}\psi|\Omega_{\rm m}\rangle$}
\label{baretadpole}
\end{center}
\end{figure}

We consider a normal vacuum $|\Omega_{\rm m=0}\rangle$ in which $\langle \Omega_{\rm m=0}|\bar{\psi}\psi|\Omega_{\rm m=0}\rangle$ is zero, and a spontaneously broken one $|\Omega_{\rm m}\rangle$ in which $\langle \Omega_{\rm m}|\bar{\psi}\psi|\Omega_{\rm m}\rangle$ is nonzero.
In the Hartree-Fock approximation we set
\begin{eqnarray}
&&\langle \Omega_{\rm m}|\left[\bar{\psi}\psi-\frac{m}{g}\right]^2|\Omega_{\rm m}\rangle=\langle \Omega_{\rm m}|\left[\bar{\psi}\psi-\frac{m}{g}\right]|\Omega_{\rm m}\rangle^2=0,
\nonumber\\
&&\langle \Omega_{\rm m}|[\bar{\psi}i\gamma^5\psi]^2|\Omega_{\rm m}\rangle=\langle \Omega_{\rm m}|\bar{\psi}i\gamma^5\psi|\Omega_{\rm m}\rangle^2=0,
\label{V6}
\end{eqnarray}
so that as shown in Fig. \ref{baretadpole} we have 
\begin{eqnarray}
\langle \Omega_{\rm m}|\bar{\psi}\psi|\Omega_{\rm m}\rangle=-i\int \frac{d^4k}{(2\pi)^4} {\rm Tr}\left[\frac{1}{\slashed{p}-m+i\epsilon}\right]=
\frac{m}{g}.
\label{V7}
\end{eqnarray}
We denote by $M$ the self-consistent value of $m$ for which this relation is  satisfied, with $M$ then obeying the gap equation
\begin{eqnarray}
-\frac{M\Lambda^2}{4\pi^2}+\frac{M^3}{4\pi^2}{\rm ln}\left(\frac{\Lambda^2}{M^2}\right)=\frac{M}{g},
\label{V8}
\end{eqnarray}
to not only admit of a nonvanishing solution for $M$, but to also  serve to define $g$ in terms of it. 

\section{Bare vertex vacuum energy}
\begin{figure}[htb]
\includegraphics[scale=0.35]{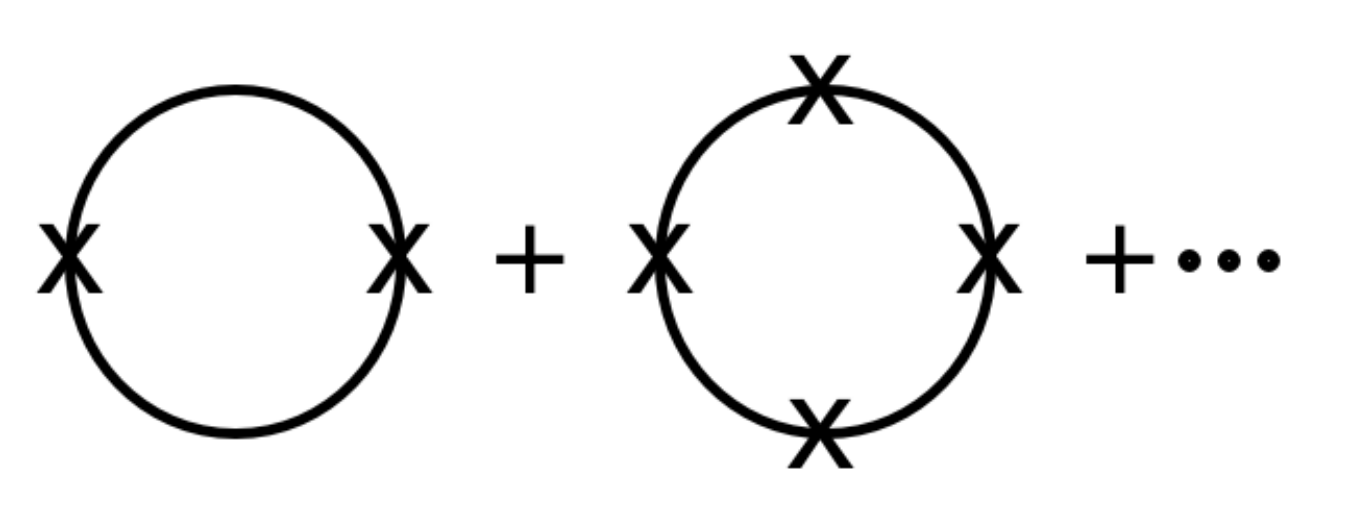}
\caption{Bare vertex vacuum energy density $\epsilon(m)$ via an infinite summation of massless graphs with zero-momentum point $m\bar{\psi}\psi$  insertions}
\label{livingwithout1}
\end{figure}
To determine which of $|\Omega_{\rm m=0}\rangle$ and $|\Omega_{\rm m}\rangle$ has the lower energy we calculate the vacuum energy by summing the infinite series of massless graphs shown in Fig. \ref{livingwithout1}.
This gives the quadratically divergent 
\begin{eqnarray}
\epsilon(m)&=&i\int \frac{d^4p}{(2\pi)^4}{\rm Tr~ln}\left[\slashed{p}-m+i\epsilon\right]-i\int \frac{d^4p}{(2\pi)^4}{\rm Tr~ln}\left[\slashed{p}+i\epsilon\right]
\nonumber\\
&=&-\frac{m^2\Lambda^2}{8\pi^2}
+\frac{m^4}{16\pi^2}{\rm ln}\left(\frac{\Lambda^2}{m^2}\right)+\frac{m^4}{32\pi^2}.
\label{V9}
\end{eqnarray}
However, now incorporating the $m^2/2g$ term we obtain 
\begin{eqnarray}
\tilde{\epsilon}(m)&=&\epsilon(m)-\frac{m^2}{2g}
=\frac{m^4}{16\pi^2}{\rm ln}\left(\frac{\Lambda^2}{m^2}\right)-\frac{m^2M^2}{8\pi^2}{\rm ln}\left(\frac{\Lambda^2}{M^2}\right)
+\frac{m^4}{32\pi^2}
\nonumber\\
&=&\frac{1}{16\pi^2}{\rm ln}\left(\frac{\Lambda^2}{M^2}\right)(m^4-2m^2M^2)+\frac{m^4}{16\pi^2}{\rm ln}\left(\frac{M^2}{m^2}\right)+\frac{m^4}{32\pi^2},
\label{V10}
\end{eqnarray}
to give an expression that is only log divergent, with an albeit cut-off dependent double-well potential emerging in the leading ${\rm ln}(\Lambda^2/M^2)$ term. With $\tilde{\epsilon}(m=M)$ being less than $\tilde{\epsilon}(m=0)$, $|\Omega_{\rm m=M}\rangle$  does indeed have lower energy than $|\Omega_{\rm m=0}\rangle$.

We note that while $\epsilon(m)$ is quadratically divergent, the half-integer spin fermion-loop generated $i\int (d^4p/(2\pi)^4)\rm Tr~ln[\slashed{p}-m+i\epsilon]$ actually has a mass-independent quartically divergent piece. In flat space it is not observable, with the $-i\int (d^4p/(2\pi)^4)\rm Tr~ln[\slashed{p}+i\epsilon]=-(i/2)\int (d^4p/(2\pi)^4)\rm Tr~ln[-p^2-i\epsilon]$ term normal ordering it away, since in flat space only energy differences are observable, and not energies themselves. Once we couple to gravity we can no longer normal order energy away and will then need gravity itself to cancel the quartic divergence. However for gravity to be able to do so  its radiative corrections would need to be under control, so that as a quantum theory it would need to be renormalizable. Conformal gravity is such a theory, and it precisely provides the needed $-(i/2)\int (d^4p/(2\pi)^4)\rm Tr~ln[-p^2-i\epsilon]$ term, with the  negative overall sign of $-(i/2)\int (d^4p/(2\pi)^4)\rm Tr~ln[-p^2-i\epsilon]$ compared to the half-integer fermion loop contribution being due to the fact that gravitons have integer spin. Since the quartic divergence is mass independent, it is cancelled on the light cone.  Below we will take care of the  remaining mass-dependent log divergence in $\tilde{\epsilon}(m)$ by dressing the point vertices in Fig. \ref{livingwithout1} with scale invariant QED interactions.

\section{Higgs-like Lagrangian}

If we make $m$ a function of $x$, i.e., we evaluate matrix elements of $\bar{\psi}\psi$ in a coherent state $|C\rangle$ \cite{Mannheim1976},  we can determine the change in the vacuum energy by noting that the vacuum to vacuum functional due to the presence of an $m(x)\bar{\psi}\psi$ term is given as $\langle\Omega(t=\infty)|\Omega(t=-\infty)\rangle=e^{iW(m(x))}$, where the associated $W(m(x))$ is of the form
\begin{align}
W(m(x))&=\displaystyle{\sum}\frac{1}{n!}\displaystyle{\int}d^4x_1...d^4x_nG^{n}_0(x_1,...,x_n)m(x_1)...m(x_n)
\nonumber\\
&=\displaystyle{\int} d^4x\left[-\epsilon(m(x))+\frac{1}{2}Z(m(x))\partial_{\mu}m(x)\partial^{\mu}m(x)+.....\right].
\label{V10a}
\end{align}
Here the $G^{n}_0(x_1,...,x_n)$ are connected Green's functions as  evaluated in the $m=0$ theory. 
$W(m(x))$ is shown graphically in Fig. \ref{livingwithout1} only with $m$ being replaced by $m(x)$ at each insertion point. The infinite summation 
gives rise to a log divergent $Z(m(x))$ 
\begin{align}
&Z(m(x))=\frac{1}{8\pi^2}\left[{\rm ln}\left(\frac{\Lambda^2}{m^2(x)}\right)-\frac{5}{3}\right],
\label{V10b}
\end{align}
and a leading log divergent effective Higgs Lagrangian of the form \cite{Eguchi1974,Mannheim1976}
\begin{align}
&{\cal{L}}_{\rm EFF}=-\epsilon(m(x))+\frac{1}{2}Z(m(x))\partial_{\mu}m(x)\partial^{\mu}m(x)+\frac{m^2(x)}{2g}
\nonumber\\
&=\frac{1}{8\pi^2}{\rm ln}\left(\frac{\Lambda^2}{M^2}\right)\bigg[
\frac{1}{2}\partial_{\mu}m(x)\partial^{\mu}m(x)
+m^2(x)M^2-\frac{1}{2}m^4(x)\bigg].~~
\label{L22}
\end{align}
With its kinetic energy term $\cal{L}_{\rm EFF}$ is recognized as the standard model  Higgs Lagrangian, except that now $m(x)$ is not an elementary quantum field, but is instead  the c-number matrix element  $m(x)=\langle C|\bar{\psi}\psi|C\rangle$.  While the  mean-field $m(x)$ cannot be an observable Higgs field since it is just a c-number matrix element, we now show that there is still an observable Higgs field  in the NJL theory, only it comes dynamically not from the mean field at all but from the residual interaction instead.

\section{The collective Goldstone and Higgs modes when the fermion is massive}

To determine bound states we first evaluate the scalar 
\begin{eqnarray}
\Pi_{\rm S}(q^2,M)
&&=-i\int\frac{d^4p}{(2\pi)^4}{\rm Tr}\left[\frac{1}{\slashed{p}-M+i\epsilon}\frac{1}{\slashed{p}+\slashed{q}-M+i\epsilon}\right]
\nonumber\\
&&=
-\frac{\Lambda^2}{4\pi^2}
+\frac{M^2}{4\pi^2}{\rm ln}\left(\frac{\Lambda^2}{M^2}\right)
+\frac{(4M^2-q^2)}{8\pi^2} +\frac{(4M^2-q^2)}{8\pi^2}{\rm ln}\left(\frac{\Lambda^2}{M^2}\right)
\nonumber\\
&&
-\frac{1}{8\pi^2}\frac{(4M^2-q^2)^{3/2}}{(-q^2)^{1/2}}
{\rm ln}\left(\frac{(4M^2-q^2)^{1/2}+(-q^2)^{1/2}}{(4M^2-q^2)^{1/2}-(-q^2)^{1/2}}\right),
\label{V15}
\end{eqnarray}
and the pseudoscalar
\begin{eqnarray}
\Pi_{\rm P}(q^2,M)&&=
-i\int\frac{d^4p}{(2\pi)^4}{\rm Tr}\left[i\gamma_5\frac{1}{\slashed{p}-M+i\epsilon}i\gamma_5\frac{1}{\slashed{p}+\slashed{q}-M+i\epsilon}\right]
\nonumber\\
&&=-\frac{\Lambda^2}{4\pi^2}
+\frac{M^2}{4\pi^2}{\rm ln}\left(\frac{\Lambda^2}{M^2}\right) 
-\frac{q^2}{8\pi^2}{\rm ln}\left(\frac{\Lambda^2}{M^2}\right) +\frac{(4M^2-q^2)}{8\pi^2}
\nonumber\\
&&
+\frac{(8M^4-8M^2q^2+q^4)}{8\pi^2 (-q^2)^{1/2}(4M^2-q^2)^{1/2}}
{\rm ln}\left(\frac{(4M^2-q^2)^{1/2}+(-q^2)^{1/2}}{(4M^2-q^2)^{1/2}-(-q^2)^{1/2}}\right).
\label{V16}
\end{eqnarray}
Iterating them in the scalar and pseudoscalar channel scattering $T$ matrices that are generated by the residual interaction gives 
\begin{align}
T_{\rm S}(q^2,M)&=g +g\Pi_{\rm S}(q^2,M)g+g\Pi_{\rm S}(q^2,M)g\Pi_{\rm S}(q^2,M)g+....=
\frac{1}{g^{-1}-\Pi_{\rm S}(q^2,M)},
\nonumber\\
T_{\rm P}(q^2,M)&=\frac{1}{g^{-1}-\Pi_{\rm P}(q^2,M)}.
\label{V17}
\end{align}
Both of these channels have poles, and near them the respective $T$ matrices behave as
\begin{align}
T_{\rm S}(q^2,M)&=\frac{R_{\rm S}^{-1}}{(q^2-4M^2)},\qquad 
T_{\rm P}(q^2,M)=\frac{R_{\rm P}^{-1}}{q^2},
\label{V17a}
\end{align}
where
\begin{eqnarray}
R_{\rm S}&=&R_{\rm P}=\frac{1}{8\pi^2}{\rm ln}\left(\frac{\Lambda^2}{M^2}\right).
\label{V18}
\end{eqnarray}
We thus dynamically generate  a massless Goldstone boson just as required of spontaneously breaking the continuous chiral symmetry. Also we generate a massive scalar Higgs boson. Since the Higgs boson is not elementary there is no Higgs boson mass hierarchy problem. Rather, the dynamically generated scalar Higgs mass is automatically finite, being naturally of order the dynamical fermion mass without any fine tuning being needed. And not only that, both its residue and that of the Goldstone boson are completely determined, though both of the residues are log divergent.

\section{Adding in scale invariance}

To eliminate the dependence on the cut off of the gap equation, of $\tilde{\epsilon}(m)$, and of $R_{\rm S}$ and $R_{\rm P}$, we need to dress the NJL point vertices so as to soften the short distance behavior of the theory. To do this we consider dressing the vertices with  
fixed point QED interactions, and follow the nonperturbative study of  Johnson, Baker and Willey \cite{Johnson1964,Johnson1967}, a study that  predated the development of the renormalization group, and that was subsequently recast in the language of the renormalization group in \cite{Adler1971}. At such a fixed point the inverse fermion propagator behaves as 
\begin{align}
{S}^{-1}(p)= \slashed{p}-m\left(\frac{-p^2-i\epsilon}{\mu^2}\right)^{\gamma_{\theta}(\alpha)/2}, 
\label{L12a}
\end{align}
where $\gamma_{\theta}(\alpha)$ is the anomalous dimension of the normal-ordered $\theta$$=:$$\bar{\psi}\psi$$:$, $\alpha$ is the point at which the renormalization group beta function $\beta(\alpha)$ vanishes, and $\mu^2$ is a renormalization group subtraction point. At such a fixed point the theory becomes scale invariant  
with a Wilson expansion of the form
\begin{eqnarray}
T(\psi(x)\bar{\psi}(0))&=&\langle \Omega_{\rm m=0}|T(\psi(x)\bar{\psi}(0))|\Omega_{\rm m=0}\rangle
+(\mu^2x^2)^{\gamma_{\theta}(\alpha)/2}{\rm :}\psi(0)\bar{\psi}(0{\rm :},
\label{L11}
\end{eqnarray}
where the normal ordering is done with respect to the unbroken massless vacuum $|\Omega_{m=0}\rangle$ (viz. $:$$\psi(0)\bar{\psi}(0)$$:=$$\psi(0)\bar{\psi}(0)-\langle \Omega_{m=0}|\psi(0)\bar{\psi}(0)|\Omega_{m=0}\rangle$). If we  now take the matrix element  of the Wilson expansion in the spontaneously broken vacuum $|\Omega_{m}\rangle$, we obtain 
\begin{align}
\tilde{S}(p)=\frac{1}{\slashed{p}}+\frac{1}{m}\left(\frac{-p^2-i\epsilon}{\mu^2}\right)^{(-\gamma_{\theta}(\alpha)/2-2)},~
\tilde{S}^{-1}(p)=\slashed{p}-m\left(\frac{-p^2-i\epsilon}{\mu^2}\right)^{(-\gamma_{\theta}(\alpha)-2)/2}.
\label{L12}
\end{align}
Compatibility with the fixed point ${S}^{-1}(p)= \slashed{p}-m((-p^2-i\epsilon)/\mu^2)^{\gamma_{\theta}(\alpha)/2}$ then gives  $\gamma_{\theta}(\alpha)=-\gamma_{\theta}(\alpha)-2$, i.e.,     $\gamma_{\theta}(\alpha)=-1$ \cite{Mannheim1975}. The compatibility thus requires a unique value for $\gamma_{\theta}(\alpha)$, a value that reduces the dimension of $\bar{\psi}\psi$ from three to two, and of $(\bar{\psi}\psi)^2$ from six to four, so that quadratic divergences become logarithmic, and the four-fermion interaction becomes renormalizable to all orders in $g$ \cite{Mannheim2017a}. In this way we can eliminate the need for the NJL model cut off.  Since $\gamma_{\theta}(\alpha)$ is negative, the asymptotic behavior of the fermion propagator is $\tilde{S}(p)=1/\slashed{p}$ to thus be on the light cone, with the $(-p^2)^{(-\gamma_{\theta}(\alpha)/2-2)}=(-p^2)^{-3/2}$ self-energy term taking us off the light cone, just as we would want. Thus being at a renormalization fixed point  gives us a propagator that exhibits an intricate connection between physics on and off the light cone.

\section{Scale invariant QED coupled to the four-fermi theory at $\gamma_{\theta}(\alpha)=-1$}

To explicitly dress the NJL point vertices  we add an electromagnetic coupling of the fermion to  the NJL action to obtain 
\begin{align}
&{\cal {L}}_{\rm QED-FF}=-\frac{1}{4}F_{\mu\nu}F^{\mu\nu}+\bar{\psi}\gamma^{\mu}(i\partial_{\mu}-eA_{\mu})\psi 
-\frac{g}{2}[\bar{\psi}\psi]^2-\frac{g}{2}[\bar{\psi}i\gamma_5\psi]^2
\nonumber\\
&=-\frac{1}{4}F_{\mu\nu}F^{\mu\nu}+\bar{\psi}\gamma^{\mu}(i\partial_{\mu}-eA_{\mu})\psi 
-m\bar{\psi}\psi +\frac{m^2}{2g}
-\frac{g}{2}\left(\bar{\psi}\psi-\frac{m}{g}\right)^2-\frac{g}{2}\left(\bar{\psi}i\gamma_5\psi\right)^2
\nonumber\\
&={\cal{L}}_{\rm QED-MF}+{\cal{L}}_{\rm QED-RI},
\label{V19}
\end{align}
as broken up into mean-field and residual-interaction components.  With the QED dressing being taken to be at a fixed point with anomalous dimensions, the renormalized inverse fermion propagator is given in (\ref{L12a}), and the renormalized vertex function with a $\bar{\psi}\psi$ insertion with momentum $q_{\mu}$ is given by  \cite{Mannheim1978}
\begin{eqnarray}
 \tilde{\Gamma}_{\rm S}(p,p+q,q)&=&\left(\frac{(-p^2-i\epsilon)}{\mu^2}\frac{(-(p+q)^2-i\epsilon}{\mu^2}\right)^{\gamma_{\theta}(\alpha)/4}.
\label{V20}
\end{eqnarray}
\begin{figure}[H]
\begin{center}
\includegraphics[scale=0.35]{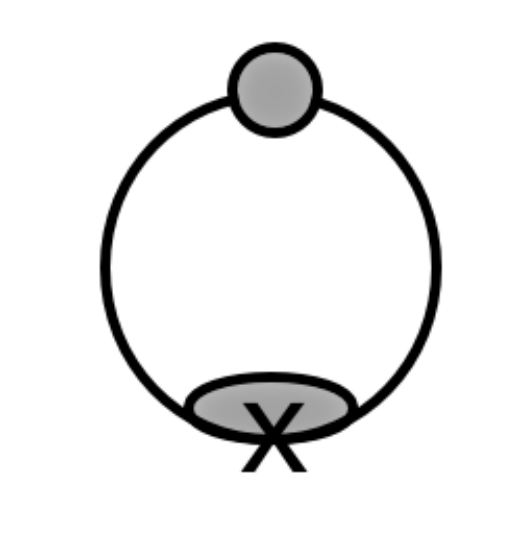}
\caption{The NJL dressed vertex  vacuum tadpole mass insertion graph for $\langle \Omega_{\rm m}|\bar{\psi}\psi|\Omega_{\rm m}\rangle$}
\label{livingwithout3}
\end{center}
\end{figure}
Dressing the vacuum tadpole graph with the propagator given in (\ref{L12a}) and the vertex given in (\ref{V20}) as in Fig.  \ref{livingwithout3} gives a gap equation of the form \cite{Mannheim1975}
\begin{eqnarray}
\langle \Omega_m|\bar{\psi}\psi|\Omega_m\rangle&=&
-\frac{m\mu^2}{4\pi^2}{\rm ln}\left(\frac{\Lambda^2}{m\mu}
\right)=\frac{m}{g} 
\label{V21}
\end{eqnarray}
when $\gamma_{\theta}(\alpha)=-1$,
with solution $M$ that obeys 
\begin{eqnarray}
-\frac{\mu^2}{4\pi^2}{\rm ln}\left(\frac{\Lambda^2}{M\mu}
\right)=\frac{1}{g},~~~~M=\frac{\Lambda^2}{\mu}\exp\left(\frac{4\pi^2}{\mu^2g}\right). 
\label{V22}
\end{eqnarray}
The gap equation gives $-g \sim 1/{\rm ln}\Lambda^2$. Thus $g$ is negative, i.e., attractive, becoming very small as $\Lambda \rightarrow \infty$, with a superconductivity BCS-type essential singularity in $M$ at $g=0$. Hence dynamical symmetry breaking can occur with weak coupling.

\section{Dressed  vertex  vacuum energy}
\begin{figure}[H]
\begin{center}
\includegraphics[scale=0.35]{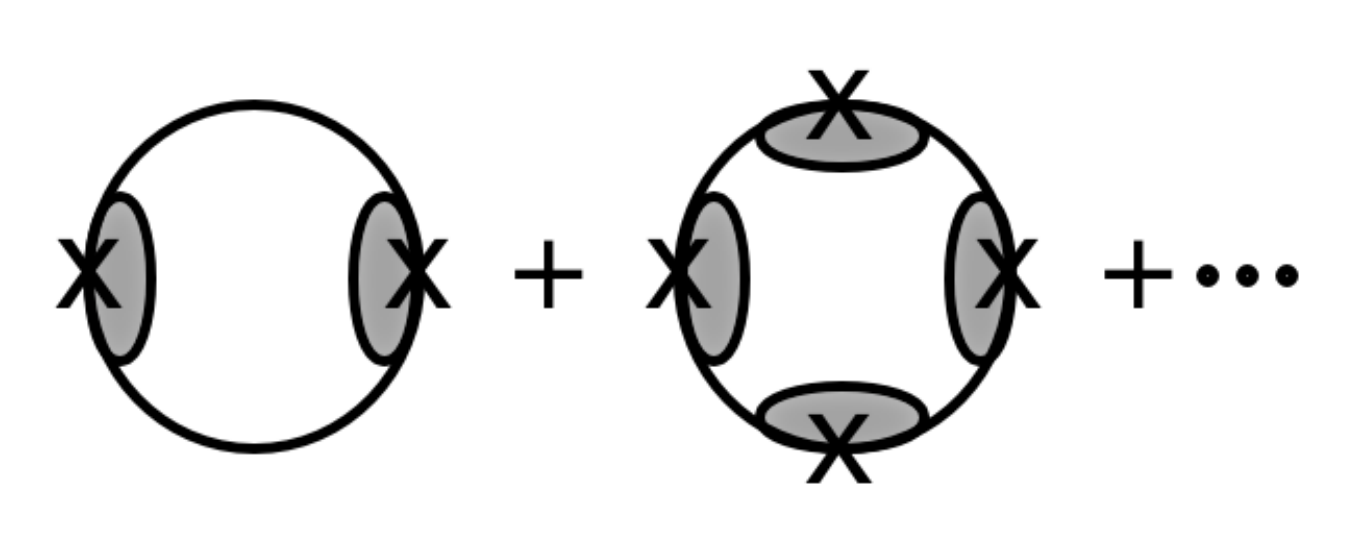}
\caption{Dressed  vertex  vacuum energy density $\epsilon(m)$ via an infinite summation of massless graphs with  dressed  zero-momentum $m\bar{\psi}\psi$  vertex insertions}
\label{livingwithout2}
\end{center}
\end{figure}
\begin{figure}[H]
\includegraphics[scale=0.35]{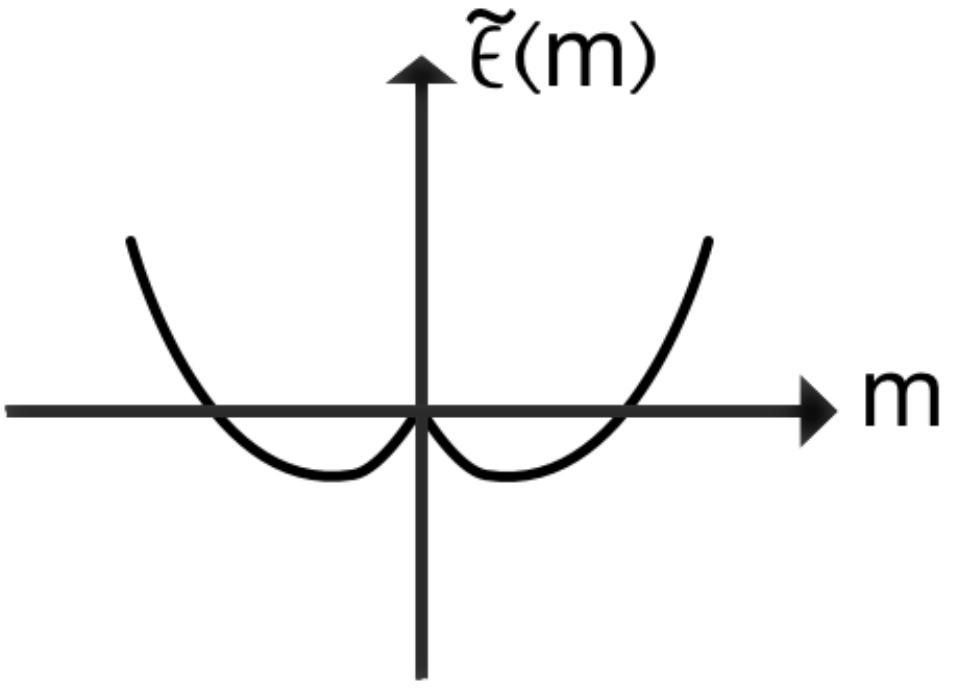}
\caption{Dressed  vertex  dynamically generated double-well potential for the renormalized vacuum energy density when $\gamma_{\theta}(\alpha)=-1$}
\label{livingwithout4}
\end{figure}

Dressing the vacuum energy as in Fig.  \ref{livingwithout2} gives 
\begin{align}
\epsilon(m)&=i\int \frac{d^4p}{(2\pi)^4}\left[{\rm Tr~ln}\tilde{S}^{-1}(p)-{\rm Tr~ln}\slashed{p}\right]
=-\frac{m^2\mu^2}{8\pi^2}\left[ {\rm ln}\left(\frac{\Lambda^2}{m\mu}
\right)+\frac{1}{2}\right],
\label{V23}
\end{align}
when $\gamma_{\theta}(\alpha)=-1$,
with $\epsilon(m)$ being  only log divergent. On including the  $m^2/2g$ term we obtain the completely finite \cite{Mannheim1975}
\begin{eqnarray}
\tilde{\epsilon}(m)=\epsilon(m)-\frac{m^2}{2g}=\frac{m^2\mu^2}{16\pi^2}\left[{\rm ln}\left(\frac{m^2}{M^2}\right)-1\right].
\label{V24}
\end{eqnarray}
We recognize $\tilde{\epsilon}(m)$ as the double-well potential shown in Fig. \ref{livingwithout4}, one dynamically induced.
We thus see the power of dynamical symmetry breaking. It reduces divergences. Moreover, since $m^2/2g$ is a cosmological term, dynamical symmetry breaking has a control over the cosmological constant problem that an elementary Higgs field potential does not. As we show below, when coupled to conformal gravity (as needed for the quartic divergence in the vacuum energy as discussed above), the cosmological constant problem is completely solved.

\section{Higgs-like Lagrangian}
\begin{figure}[H]
\begin{center}
\includegraphics[scale=0.35]{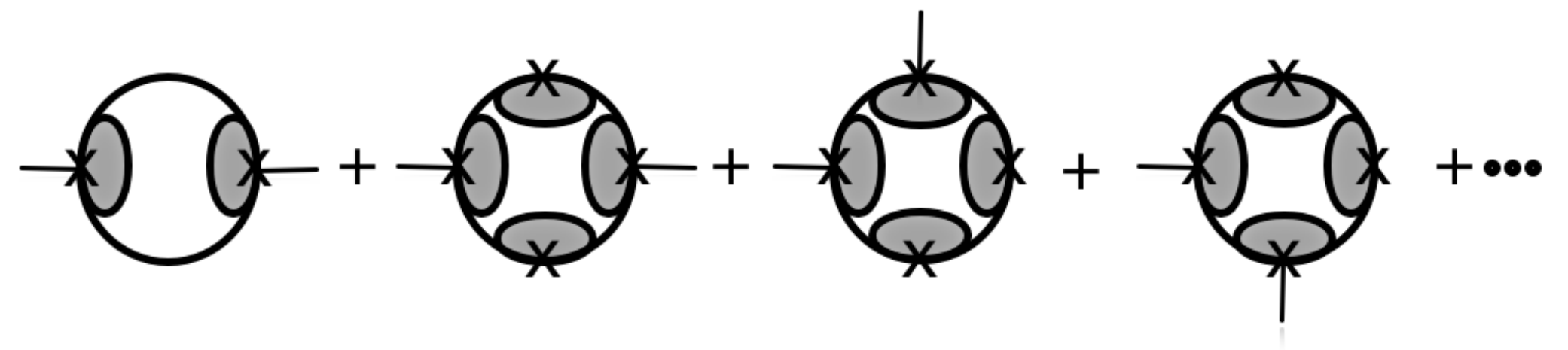}
\caption{The infinite series of massless graphs with two external insertions each carrying momentum $q^{\mu}$  (shown as straight external lines) and zero-momentum $\bar{\psi}\psi$ insertions (shown as crosses) needed to determine  $\Pi_{\rm S}(q^2,m(x))$ (two external lines coupling as scalars) and $\Pi_{\rm P}(q^2,m(x))$ (two external lines coupling as pseudoscalars)}
\label{livingwithout6}
\end{center}
\end{figure}
To obtain an effective Lagrangian when $\gamma_{\theta}(\alpha)=-1$ we sum the infinite set of graphs given in Fig. \ref{livingwithout6}, to obtain \cite{Mannheim1978} the completely finite
\begin{eqnarray}
{\cal{L}}_{\rm EFF}&&=-\epsilon(m(x))+\frac{1}{2}Z(m(x))\partial_{\mu}m(x)\partial^{\mu}m(x)+\frac{m^2}{2g}
\nonumber\\
&&=-\frac{m^2(x)\mu^2}{16\pi^2}\left[{\rm ln}\left(\frac{m^2(x)}{M^2}\right)-1\right]
+\frac{3\mu}{256\pi m(x)}\partial_{\mu}m(x)\partial^{\mu}m(x)+....,
\label{L56}
\end{eqnarray}
where there are higher derivative terms along with the kinetic energy. We recognize (\ref{L56}) as  effective Higgs-like Lagrangian.

\section{The collective tachyon modes when the fermion is massless}
In order to appreciate why at $\gamma_{\theta}(\alpha)=-1$ there have to be massless Goldstone and massive Higgs bosons in the massive fermion  scattering $T$ matrix, we first evaluate the scattering $T$ matrix using dressed vertices but while  keeping the fermion massless. 
\begin{figure}[H]
\begin{center}
\includegraphics[scale=0.45]{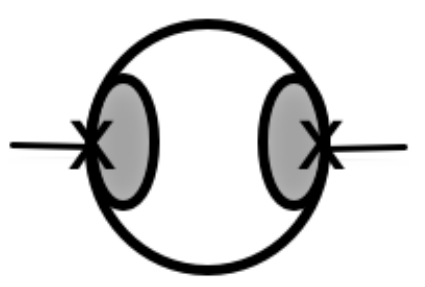}
\caption{Dressed  vertex  interaction graph needed for $\Pi_{\rm S}(q^2,m=0)$ and $\Pi_{\rm P}(q^2,m=0)$}
\label{livingwithout6cropped}
\end{center}
\end{figure}
With $\Pi_{\rm S}(q^2,m=0)=\Pi_{\rm P}(q^2,m=0$) being given in Fig. \ref{livingwithout6cropped}, they evaluate to 
\begin{eqnarray}
&&\Pi_{\rm S}(q^2,m=0)=\Pi_{\rm P}(q^2,m=0)
=-\frac{\mu^2}{4\pi^2}\bigg[{\rm ln}\left(\frac{\Lambda^2}{(-q^2)}\right)-3+4~{\rm ln}2\bigg].
\label{V25}
\end{eqnarray}
The associated $T$ matrices are thus of the form 
\begin{eqnarray}
T_{\rm S}(q^2,m=0)&=&\frac{g}{1-g\Pi_{\rm S}(q^2,m=0)}=\frac{1}{g^{-1}-\Pi_{\rm S}(q^2,m=0)},
\nonumber\\
T_{\rm P}(q^2,m=0)&=&\frac{g}{1-g\Pi_{\rm P}(q^2,m=0)}=\frac{1}{g^{-1}-\Pi_{\rm P}(q^2,m=0)},
\label{V26}
\end{eqnarray}
and both have a pole at 
\begin{eqnarray}
q^2=-M\mu e^{4{\rm ln}2-3}=-0.797M\mu,
\label{V27}
\end{eqnarray}
near which the $m=0$ $T$ matrices behave as 
\begin{eqnarray}
T_{\rm S}(q^2,m=0)=T_{\rm P}(q^2,m=0)=\frac{31.448M\mu}{(q^2+0.797M\mu)}.
\label{V28}
\end{eqnarray}
That the poles are degenerate is because with massless fermions the chiral symmetry is unbroken. That the poles are both in the tachyonic spacelike region means that the massless vacuum is unstable (just like being at the local maximum in a double-well potential). Thus before looking to see what happens when the fermion is massive, we already know that the theory cannot support any massless fermion at all. So we now consider what happens when the fermion does acquire a mass. 

\section{The collective Goldstone mode when the fermion is massive}

Summing Fig. \ref{livingwithout6} with the two external lines coupling as pseudoscalars, with a dressed massive fermion propagator and dressed  vertices, the pseudoscalar  $\Pi_{\rm P}(q^2=0,m)$ is given by \cite{Mannheim2017c}

\begin{eqnarray}
\Pi_{\rm P}(q^2=0,m)&=&-4i\mu^2\int \frac{d^4p}{(2\pi)^4}\frac{(p^2)(-p^2)-M^2\mu^2}{((p^2+i\epsilon)^2+M^2\mu^2)^2}
\nonumber\\
&=&4i\mu^2\int \frac{d^4p}{(2\pi)^4}\frac{1}{(p^2+i\epsilon)^2+M^2\mu^2}=\frac{1}{g}.
\label{V29}
\end{eqnarray}
when $m=M$ and $M$ satisfies (\ref{V22}).
This puts a massless Goldstone boson in the pseudoscalar $T$ matrix channel of the form 
\begin{eqnarray}
T_{\rm P}(q^2,M)=\frac{128\pi M}{7\mu q^2}=\frac{57.446 M}{\mu q^2}
\label{V30}
\end{eqnarray}
near $q^2=0$. Not only is the residue completely calculable, unlike the point vertex case given in (\ref{V18}), the residue  is completely finite.

\section{The collective Higgs mode when the fermion is massive}

The scalar channel calculation repeats as in the pseudoscalar channel, and leads \cite{Mannheim2017c} to a dynamically massive Higgs boson with 
\begin{eqnarray}
q_0(\rm Higgs)&=&(1.480-0.017i)(M\mu)^{1/2},\qquad
q^2(\rm Higgs)=(2.189-0.051i)M\mu.
\label{V31}
\end{eqnarray}
If we set  $\mu=M$ these expressions reduce to 
\begin{eqnarray}
q_0(\rm Higgs)&=&(1.480-0.017i)M,\qquad
q^2(\rm Higgs)=(2.189-0.051i)M^2.
\label{V32}
\end{eqnarray}
Near the Higgs pole the scalar channel $T$ matrix is found to behave as 
\begin{align}
T_{\rm S}(q^2,M)=\frac{46.141+1.030i}{q^2-2.189M^2+0.051i M^2},
\label{V32a}
\end {align}
to also have a finite, completely calculable residue.
The Higgs boson mass is close to the dynamical fermion mass, but above the fermion-antifermion threshold, to thus have a width. In a double-well elementary Higgs field theory the Higgs mass is real, since it is given by the second derivative of a real Higgs potential at its minimum. The width can thus be used to distinguish an elementary Higgs boson from a dynamical one.

In regard to the cancellation of the vacuum energy divergence we also note that we can  cancel the quartic divergence in the matter sector vacuum energy by  a quartic divergence in the gravity sector, provided the gravity sector is itself renormalizable, as is the case with conformal gravity. Together with dynamical mass generation this then controls the cosmological constant \cite{Mannheim2017}. Specifically, the vacuum energy  of a massive  fermion has a quartic divergence, a quadratic divergence, a logarithmic divergence, and a finite part. Conformal gravity takes care of the quartic divergence. Critical scaling in the matter sector and the reduction in the dimension of $\bar{\psi}\psi$ from three to two reduces the quadratic divergence to logarithmic, and the mean field induced $-m^2/2g$ term when mass is generated dynamically then takes care of the resulting logarithmic divergence, leaving the vacuum energy completely finite. In addition, by being dynamically generated the associated Higgs boson has no hierarchy problem.

\section{Fitting the accelerating universe data}
\begin{figure}[H]
\begin{center}
\includegraphics[height=3.2in,width=5.0in]{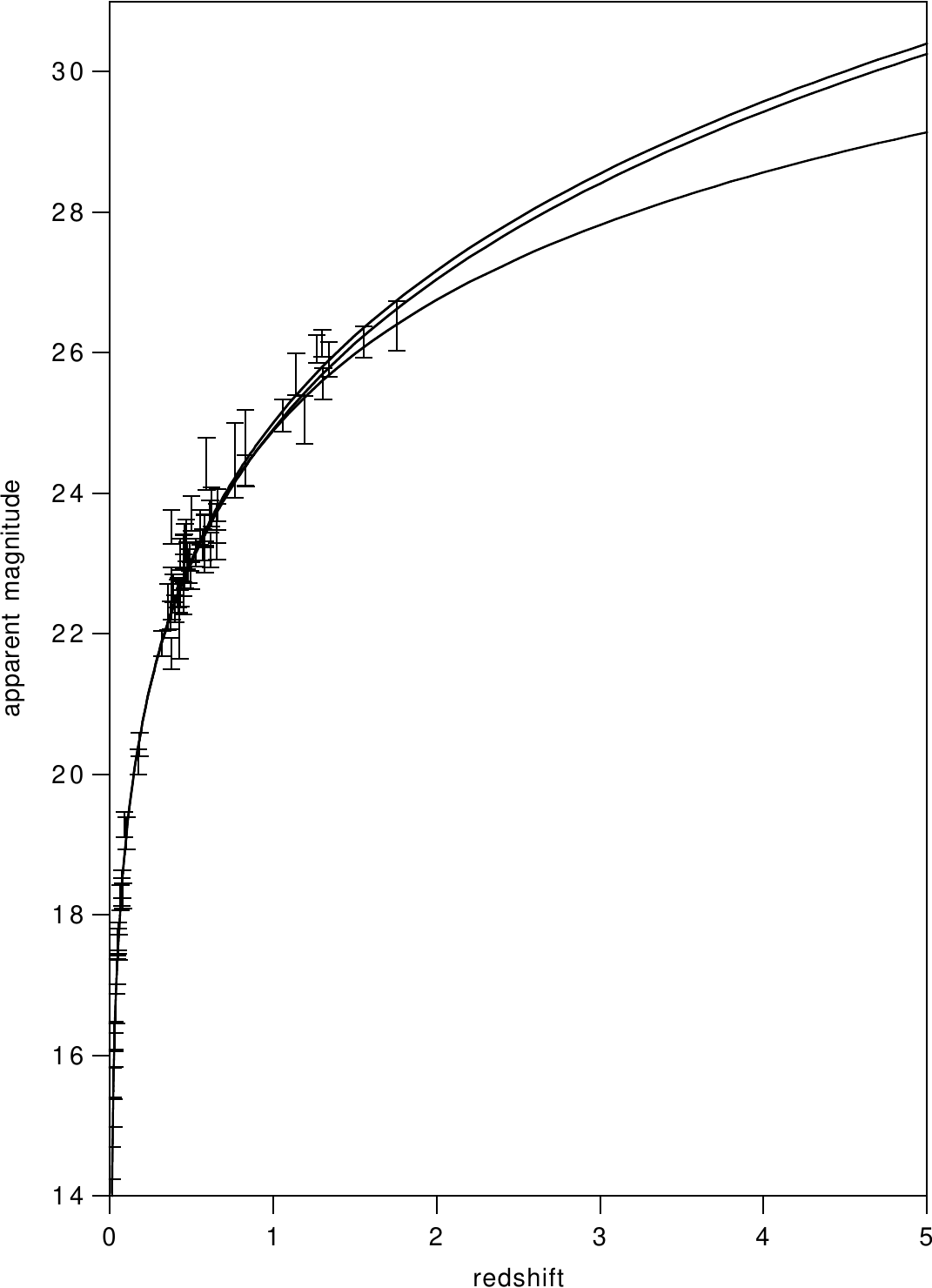}
\caption{Apparent magnitude versus redshift Hubble plot expectations for $q_0=-0.37$ (highest curve) and
$q_0=0$ (middle curve) conformal gravity and for
$\Omega_{M}(t_0)=0.3$,
$\Omega_{\Lambda}(t_0)=0.7$ standard gravity (lowest curve)}
\label{Graph2bw}
\end{center}
\end{figure}

Not only can we obtain a finite cosmological constant if conformal gravity is coupled to dynamical symmetry breaking, we can even get one whose contribution to cosmic expansion  is so under control that it can fit the cosmological Hubble plot data without any of the large amount of fine tuning that is needed in standard $\Lambda$CDM. Specifically, we can show \cite{Mannheim2006,Mannheim2017} that in a  conformal gravity Robertson-Walker cosmology the current era value $q_0$ of the deceleration parameter has to lie in the very narrow  range $-1\leq q_0\leq 0$ no matter what the value of the parameters in the model or how big they might be. With $H_0$ being the current value of the Hubble parameter, the associated luminosity function $d_L(z)$ as a function of redshift $z$ is given by \cite{Mannheim2006}: 
\begin{align} 
d_L(z)=-\frac{c}{H_0}\frac{(1+z)^2}{q_0}\left(1-\left[1+q_0-
\frac{q_0}{(1+z)^2}\right]^{1/2}\right).~~
\label{L107}
\end{align}
As shown in Fig. \ref{Graph2bw} with the form for the apparent magnitude associated with $d_L(z)$, the conformal gravity fit to the accelerating universe data of \cite{Riess1998,Perlmutter1999} is of a quality comparable to that of the standard model dark matter dark energy fit, with the best conformal gravity fit value for $q_0$  being  given by  $q_0=-0.37$ \cite{Mannheim2006}, a value that non-trivially is right in the required $-1\leq q_0\leq 0$ range. The cosmological constant term is thus indeed under control.

\section{Light-front axial-vector Ward identity}

Since the above discussion of dynamical symmetry breaking only involves Feynman diagrams, the outcome is the same in both instant-time quantization and light-front quantization, though in the light-front case we need to include the circle at infinity contributions to the bare and dressed tadpole graphs given in Fig. \ref{baretadpole} and Fig. \ref{livingwithout3}. While the same spontaneously broken symmetry outcome must also occur in the axial vector Ward identity, the way that it does so in the  light-front case is somewhat different from the way it does so  in the instant-time case. This is because of the role played by the light-front bad fermions.

To see the issues involved, following \cite{Mannheim2021} we analyze the components of  the axial-vector current  $A^{\mu}=\bar{\psi}\gamma^{\mu}\gamma^5\psi$. In light-front components $ A^+=2\psi^{\dagger}_{(+)}\gamma^5\psi_{(+)}$ is written in terms of good fermions alone,  $ A^-=2\psi^{\dagger}_{(-)}\gamma^5\psi_{(-)}$ is written in terms of bad fermions alone, and  $ A^1=\psi^{\dagger}\gamma^0\gamma^{1}\gamma^5\psi$ and $A^2=\psi^{\dagger}\gamma^0\gamma^{2}\gamma^5\psi$ contain both good and bad fermions. We take the axial-vector current to be conserved so that $\partial_+A^++\partial_-A^-+\partial_1A^1+\partial_2A^2=0$. While the axial charge $Q{^5}=(1/2)\int dx^-dx^1dx^2A^+= \int dx^-dx^1dx^2\psi^{\dagger}_{(+)}\gamma^5\psi_{(+)}$ only contains good fermions, its light-front time derivative $\partial_+Q^5=-(1/2)\int dx^-dx^1dx^2(\partial_-A^-+\partial_1A^1+\partial_2A^2)$ involves both good and bad fermions. Since $\psi_{(-)}$ obeys  the nonlocal $\psi_{(-)}=-(i/2)(\partial_-)^{-1}[-i\gamma^0(\gamma^1\partial_1+\gamma^2\partial_2)+m\gamma^0]\psi_{(+)}$ given in (\ref{LC4.4}), to secure the light-front time independence of $Q^5$ requires that the fermion fields be more convergent asymptotically than in the instant-time case. In addition, the scalar and pseudoscalar fermion bilinears  are of  the form 
\begin{eqnarray}
\bar{\psi}\psi=\psi_{(+)}^{\dagger}\gamma^0\psi_{(-)}+\psi_{(-)}^{\dagger}\gamma^0\psi_{(+)},\quad \bar{\psi}i\gamma^5\psi=\psi_{(+)}^{\dagger}i\gamma^0\gamma^5\psi_{(-)}+\psi_{(-)}^{\dagger}i\gamma^0\gamma^5\psi_{(+)},
\label{B10a}
\end{eqnarray}
and thus they both contain both good and bad fermions.

Noting that generically we have 
\begin{eqnarray}
A^{\dagger}BC^{\dagger}D-C^{\dagger}DA^{\dagger}B&=
A^{\dagger}(BC^{\dagger}+C^{\dagger}B)D-C^{\dagger}(DA^{\dagger}+A^{\dagger}D)B
\nonumber\\
&
-(A^{\dagger}C^{\dagger}+C^{\dagger}A^{\dagger})BD+C^{\dagger}A^{\dagger}(BD+DB),
\label{C5}
\end{eqnarray}
on using the equal light-front time anticommutators given earlier we obtain
\begin{align}
&[Q^5,\bar{\psi}(x)i\gamma^5\psi(x)]=
\nonumber\\
&\int dy^-dy^1dy^2[\psi^{\dagger}_{(+)}(y)\gamma^5\psi_{(+)}(y),
i\psi_{(+)}^{\dagger}(x)\gamma^0\gamma^5\psi_{(-)}(x)+i\psi_{(-)}^{\dagger}(x)\gamma^0\gamma^5\psi_{(+)}(x)]
\nonumber\\
&=i\psi_{(+)}^{\dagger}(x)\gamma^0\psi_{(-)}(x)+i\psi_{(-)}^{\dagger}(x)\gamma^0\psi_{(+)}(x)=i\bar{\psi}(x)\psi(x).
\label{C6}
\end{align}
Thus despite the presence of both good and bad fermions, they organize themselves to give  $ [Q^5,\bar{\psi}(x)i\gamma^5\psi(x)]=i\bar{\psi}(x)\psi(x)$, i.e., to give precisely the same form as in the instant-time case.

We introduce the vacuum matrix element of the light-front time-ordered product $\langle \Omega [|\theta(x^+)A^{\mu}(x)\bar{\psi}(0)i\gamma^5\psi(0)+\theta(-x^+)\bar{\psi}(0)i\gamma^5\psi(0)A^{\mu}(x)]|\Omega\rangle$. Since there is only one associated momentum vector in Fourier space, we can set 
\begin{align}
&\langle \Omega |[\theta(x^+)A^{\mu}(x)\bar{\psi}(0)i\gamma^5\psi(0)+\theta(-x^+)\bar{\psi}(0)i\gamma^5\psi(0)A^{\mu}(x)]|\Omega\rangle
\nonumber\\
&=\frac{1}{(2\pi)^4}\int d^4p e^{ip\cdot x}p^{\mu}F(p^2),
\label{C2}
\end{align}
where $F(p^2)$ is a scalar function. With $\partial _{\mu}A^{\mu}=0$ we apply $\partial_{\mu}$ and then  $\int d^4x$ to (\ref{C2}) to obtain 
\begin{align}
&\delta(x^+)\langle\Omega |[A^{\mu}(x),\bar{\psi}(0)i\gamma^5\psi(0)]|\Omega\rangle=\frac{i}{(2\pi)^4}\int d^4p e^{ip\cdot x}p^2F(p^2),
\label{C9}
\end{align}
\begin{align}
&i\int d^4p\delta^4(p)p^2F(p)=\langle \Omega |[Q^5(x^+=0),\bar{\psi}(0)i\gamma^5\psi(0)]|\Omega\rangle= i\langle \Omega | \bar\psi(0)\psi(0)|\Omega \rangle.
\label{C3}
\end{align}
Thus  if $\partial_{\mu}A^{\mu}=0$ and $|\Omega \rangle$ is such that $i\langle \Omega | \bar\psi(0)\psi(0)|\Omega \rangle\neq 0$, $Q^5$ must not annihilate the vacuum  and $F(p)$ must contain a pole at $p^2=0$. This then is how the Goldstone theorem is satisfied in the light-front case, with the bad fermions playing a central role.

\section{The moral of the story}

\textbf{There is a lot of interesting physics on the light cone, and even more interesting physics off it.}
\bigskip

\noindent
\textbf{Data availability}
Data sharing not applicable to this article as no datasets were generated or analyzed during the current study.

\end{document}